\documentclass{emulateapj}
\usepackage{xcolor}
\usepackage[T1]{fontenc}
\usepackage{bm}
\usepackage{multirow}
\usepackage{amsmath}

\begin{document}

\def\be{\begin{eqnarray}}
\def\ee{\end{eqnarray}}
\def\bh{{\bullet}}
\def\edd{{\rm Edd}}
\def\in{{\rm in}}
\def\mb{{\rm mb}}
\def\fb{{\rm fb}}
\def\f{{\rm f}}
\def\ms{{\rm ms}}
\def\sal{{\rm Sal}}
\def\s{{\rm S}}
\def\d{{\rm d}}
\def\J{{\bf J}}
\def\disk{{\rm disk}}
\def\obs{{\rm obs}}
\def\c{{\rm c}}
\def\bol{{\rm bol}}
\def\ks{{\rm KS}}
\def\ini{{\rm ini}}
\def\fin{{\rm fin}}
\def\cl{{\rm cl}}
\def\al{{\rm al}}
\def\crit{{\rm crit}}
\def\sg{{\rm sg}}
\def\wp{{\rm warp}}
\def\bp{{\rm BP}}
\def\yr{{\rm yr}}
\def\acc{{\rm acc}}
\def\msun{M_{\odot}}
\def\isco{{\rm ISCO}}
\def\tde{{\rm TDE}}

\def\foe{\left(\frac{f_\edd}{\eta_{0.1}}\right)}

\title{Effect of accreting tidally disrupted stars on the spin evolution of
$\sim 10^6\msun$ black holes}
\author{Xiaoxia Zhang$^{1,2,3}$, Youjun Lu$^{1,2,4\dagger}$, and Zhu Liu$^{1}$}
\affil{
~$^1$~National Astronomical Observatories, Chinese Academy
of Sciences, Beijing 100012, China; ~$^{\dagger}$luyj@nao.cas.cn\\
~$^2$~School of Astronomy and Space Sciences, University of 
Chinese Academy of Sciences, No. 19A Yuquan Road, Beijing 100049, China \\
~$^3$~Department of Astronomy, Xiamen University, Xiamen, Fujian, 361005, China\\
~$^4$~CAS Key laboratory for computational Astrophysics, National Astronomical 
Observatories, Chinese Academy of Sciences, Beijing, 100012, China
}

\begin{abstract}
Accretion of tidally disrupted stars (TDSs) is expected to contribute 
significantly to the growth of massive black holes (MBHs) with mass 
$\sim 10^6\msun$ in galactic centers.  In this paper, we quantitatively 
investigate the effect of the TDS accretion on the spin evolution of these
relatively small MBHs, by also considering the accretion of gas-clouds 
with (many) chaotic episodes.  We find that the accretion of TDSs can 
play an important role or even a dominant role in shaping the spin
distribution of $\sim 10^6\msun$ MBHs, depending on the contribution
fraction ($f_\tde$) of the TDS accretion to the MBH growth. If $f_\tde$ 
is as large as $\ga 0.9$, most $\sim 10^6\msun$ MBHs have low spins 
($|a| \la0.3$); if $f_\tde$ is as small as $\la 0.1$, most $\sim 10^6\msun$ 
MBHs have high spins ($|a| \ga0.7$). We also find that (1) the fraction of
highly spinning $\sim10^6\msun$ MBHs in the TDS accretion states is 
smaller than that in the gas-cloud (AGN) accretion states, which is a 
consequence of more rapid spin decrease during the period of consecutive 
TDS accretion than the spin increase during the AGN periods when the
spin is large; (2) the fraction of retrograde spin accretion in the TDS
accretion states is almost the same as that of prograde spin accretion,
while it is negligible in the gas-cloud (AGN) accretion states.
Current scarce sample of AGNs ($\sim10^6\msun$) with spin 
measurements hints an insignificant contribution from TDS accretion 
to MBH growth. Future measurements on spins of $\sim10^6\msun$ 
MBHs may provide stronger constraints on the importance of both 
AGN and TDS accretion states in their growth history. 
\end{abstract}
%

\section{Introduction}
\label{sec:Intro}

Massive black holes (MBHs) exist almost ubiquitously in galactic
centers \citep[e.g.,][]{KH13}, which are believed to be fully specified 
by their masses and spins \citep{Kerr}. It has been demonstrated 
that the spin distribution of MBHs may contain important information 
about the MBH formation and assembly history and thus may be 
used to put constraints on their growth history \citep[e.g.,][]{vol05, 
KP06, Volonteri07, ber08, KPH08, per09, Sadowski11, Volonteri13, 
Dotti13, ses14}, in addition to the mass distribution and the AGN 
luminosity function \citep[e.g.,][]{Soltan82, YT02, Marconi04, YL04, 
YL08, Shankar09}. 

The accretion modes that MBHs experienced play a leading role in
determining the MBH spin evolution \citep[e.g.,][]{Volonteri07}.
Coherent accretion, whether via thick disk or the standard thin disk,
leads to rapid spin-up of the central MBH, while ``chaotic'' 
accretion, composed of many episodes with randomly distributed 
disk orientations, may lead to spin-down of a rapidly rotating MBH. 
Therefore, the spins of most active MBHs should be close
to $1$ if their growth is dominated by coherent accretion, and they
may be close to $0$ if their growth is dominated by chaotic accretion
with sufficiently small disk mass in each accretion episode. 

Currently there are more than two dozen MBHs in AGNs having spin
estimations, mainly by using the profile of Fe K$\alpha$ line and the
X-ray reflection spectroscopy, and most of those MBHs are rapidly
rotating with spin $\ga 0.8$  \citep[e.g.,][]{BR06, Brenneman13,
Reynolds14, vas16}. High radiative efficiencies inferred for some 
QSOs also suggest rapidly spinning MBHs in those QSOs 
\citep[e.g.,][]{2014ApJ...789L...9T, 2016MNRAS.460..212C}. 
However, \citet{Liuetal15} recently showed that those MBHs in 
narrow-line Seyfert 1s (NLS1s), typically with mass $\sim10^6\msun$, 
may not be spinning very fast in general according to the profile of the 
broad relativistic Fe K$\alpha$ line found in the stacked spectrum of a 
large number of NLS1s. It is not clear whether the apparently low spins 
of those MBHs in NLS1s are due to their different growth histories.

Tidal disruption of stars was first predicted as an exotic phenomenon
inevitably resulting from the existence of MBHs in the centers of galaxies 
\citep[e.g.,][]{Hills75, Rees88}, and later confirmed by observations 
\citep[e.g., as summarized in][]{Komossa15}. 
The predicted rate of TDEs is about $10^{-4}-10^{-3}\,{\rm yr}^{-1}$ 
for small galaxies with $\sim10^{6}\msun$ MBHs and $10^{-6}-
10^{-5}\,{\rm yr}^{-1}$ for big galaxies with $\gtrsim 10^7\msun$ 
MBHs \citep[e.g.,][]{Magorrian99, WangMerritt04, Kesden12}. By 
using TDEs found in various surveys, such as the ROSAT all-sky 
survey, the XMM-Newton slew survey, the Galaxy Evolution Explorer 
(GALEX) survey, and the Sloan Digital Sky Survey (SDSS), the TDE 
rate has also been frequently estimated and the value is generally 
$\sim 10^{-5}\ {\rm yr}^{-1}$ per galaxy \citep[e.g.,][]{donley02, 
esquej08, gazari08, vanVelzen14}. The discrepancy between the 
theoretical and observational rates could be caused by various 
factors such as selection effect, small number statistics, and dust 
extinction \citep[see][]{stone16}. If small galaxies do have a higher 
TDE rate as suggested by theoretical studies \citep[e.g.,][]{WangMerritt04}, 
the accretion of TDSs over the cosmic time would dominate or contribute 
significantly to the mass growth of $\sim 10^{6}\msun$ MBHs \citep[see, 
e.g.,][]{ML06}, though it contributes little to the growth of $\ga10^7\msun$ 
MBHs as the TDE rate for those MBHs is much smaller. 

Accretion of TDSs may also affect the spin evolution of MBHs, 
especially the small ones with mass $\sim10^6\msun$, in addition 
to the AGN accretion modes mentioned above. \citet{Volonteri07} 
have pointed out that the significance of accreting TDSs in the 
growth histories of $\sim 10^6\msun$ MBHs may lead to the spin
distribution of those MBHs significantly different from that for MBHs 
with mass substantially larger than $10^7\msun$. In this paper, we 
quantitatively investigate the effect of accreting TDSs on the spin 
evolution of $\sim 10^6\msun$ MBHs, whether this effect could lead 
to a spin distribution of those AGNs with $\sim10^6\msun$ MBHs 
significantly different from that of $>10^7 \msun$ MBHs, and whether 
it can be used to constrain the significance of TDS accretion to the 
growth of those MBHs.

This paper is organized as follows. In Section~\ref{sec:spinevol}, we
consider an evolution model for MBHs with mass $\sim
10^6\msun$\footnote{The effect of mergers is not considered in this
paper as those $\sim 10^6\msun$ MBHs, mainly in the bulges of spiral
galaxies, probably did not experience a significant number  of major
mergers in their assembly histories.}, which grow up via the accretion
of numerous tidally disrupted stars (TDSs) and a number of gas-clouds 
on orbits with random orientations relative to the MBH equatorial plane. 
More  detailed physics about the TDS and gas-cloud accretion are 
considered here compared with previous analytic studies on TDS effect 
on MBH spin evolution in \citet{vol05} and chaotic gas-cloud accretion in 
\citet{KPH08}.
Our results on the spin distribution of MBHs in both the gas-cloud (AGN) 
accretion episodes and the TDS accretion episodes obtained from some 
example models are presented in Section~\ref{sec:results}. According to 
those results, we obtain some implications on the importance of the TDS 
accretion to the growth of $\sim 10^6\msun$ MBHs from current spin 
measurements in in Section~\ref{sec:implication}. Conclusions and
discussions are given in Section~\ref{sec:conclusions}.

\section{Spin evolution of $\sim 10^6\msun$ MBHs}
\label{sec:spinevol}
To investigate the effect of accreting TDSs on the spin evolution of 
$\sim 10^6 \msun$ MBHs, we assume an initial mass of $10^5\msun$ 
and a final masses of $10^{6.5}\msun$ for those MBHs\footnote{We simply choose such a mass range as our goal in the present paper is to investigate how accretion of TDSs influences the spin evolution of $\sim10^6\msun$ MBHs. Adopting a smaller initial MBH mass does not affect our results; adopting a substantially larger final MBH mass ($>10^7\msun$) does affect our results, but we have argued that the contribution from TDS accretion to those is insignificant.}, and they grow 
up by accreting (1) TDSs and (2) giant gas-clouds falling in. 
Since the plug-in orbits of TDSs or the infalling gas-cloud may be 
randomly oriented, the accretion disks can be either on the MBH 
equatorial plane or inclined to it. Therefore, the mass growth and 
spin evolution of the MBH are generally controlled by the amount 
of material falling into the MBH from the inner disk boundary and 
the torque exerted by the (inclined) disk on the MBH spin. In general, 
the MBH mass growth rate is determined by \citep[e.g., see][]{Thorne74}
\begin{equation}
\frac{dM_\bullet}{dt} = f_\edd \frac{E_{\rm in} }{1-E_{\rm in}}
\frac{M_\bullet}{t_{\rm Edd}}, 
\label{eq:dmdt}
\end{equation}
where $f_\edd$ is the Eddington ratio, $E_{\rm in}$ is the specific 
energy of the accreted material at the disk inner boundary $r_{\rm in}$ 
which may be coincident with the innermost stable circular orbit (ISCO), 
and $t_{\rm Edd} =4.5\times 10^8\ \textrm{yr}$ is the Eddington timescale. 
The MBH angular momentum ($\mathbf{J}_\bullet$) evolution is controlled by
\begin{equation}
\frac{d\mathbf{J}_\bullet}{dt} = \dot{M} r_{\rm g} c L_{\rm in}
\hat{\mathbf{l}} + \frac{4\pi G}{c^2} \int_{\disk} \frac{ \mathbf{L}
\times \mathbf{J}_{\bullet}} {r^2\cdot  r_ {\rm g} } dr, 
\label{eq:spinevol}
\end{equation}
where $\dot{M}$ is the accretion rate at $r_{\rm in}$, $L_{\rm in}$ is 
the specific 
angular momentum
at $r_{\rm in}$, $\mathbf{L}$ is the angular momentum vector of disk per unit 
area, $\hat{\mathbf{l}}$ is a unit vector parallel to $\mathbf{L}(r_{\rm in})$,   
and $r$ is in unit of gravitational radius $r_{\rm g}$ ($= GM_{\bullet} /c^2$ 
with $G$ the gravitational constant and $c$ the speed of light; see also 
\citealt{per09, Dotti13} for Eq.~(\ref{eq:spinevol})). The first term at the r.h.s. 
of Equation~(\ref{eq:spinevol}) represents the contribution from the angular 
momentum brought in by material falling from the disk inner boundary, and 
the second term at the r.h.s. of Equation~(\ref{eq:spinevol}) represents the 
coupling between the MBH spin and the disk angular momentum 
i.e., the Lense-Thirring (LT) precession \citep{1918PhyZ...19..156L}. 
This leads to a warped accretion disk, with the inner region bent to 
the MBH equatorial plane and the outer region keeping the original 
misalignment \citep{bar75}, and may further lead to a gradual alignment 
(or anti-alignment) of the MBH and disk angular momenta. If the initial 
angle $\beta$ between ${\bf J}_{\rm disk}$ and ${\bf J}_\bullet$ satisfies 
the condition $\cos\beta <-J_\disk/2J_\bh$, then the system will finally 
reach a stable counter-alignment configuration, and otherwise the 
alignment will be the case \citep{kin05}. 

If the disk is on the equatorial plane of the MBH, then the second 
term vanishes, and combining Equations~(\ref{eq:dmdt}), 
Equations~(\ref{eq:spinevol}) is reduced to \citep[e.g.,][]{Thorne74, 
Barausse12, Dotti13}
\begin{equation}
\frac{d a}{dt} = \left(L_{\rm in} - 2 a E_{\rm in} \right)
\frac{f_\edd} {(1-E_{\rm in}) t_{\rm Edd}}, 
\label{eq:dadt}
\end{equation}
where $E_{\rm in}$ and $L_{\rm in}$ are values at ISCO of 
equatorial disk, and $a$ is the dimensionless spin parameter 
with $|a| = \frac{c \left| {\bf J}_\bullet\right|} {G M^2_\bullet}$. 
Throughout this paper, $a$ is positive if the disk is co-rotating 
around the MBH, and negative if counter-rotating around the MBH. 

In this paper, we consider the accretion of TDSs and gas-clouds 
separately and ignore the possibility of accreting TDSs during the 
gas-cloud (AGN) accretion episodes. The main reasons are as follows.
First, if a star is tidally disrupted during a gas-cloud (AGN) accretion 
episode, the star before its disruption or the tidal debris after the 
disruption may ground down quickly due to its interaction with the 
accretion disk and become part of the disk \citep[cf.,][]{Art93, MK05}. 
Second, currently it is still not clear what exactly the signal is from a 
TDE occurred in the gas-cloud (AGN) accretion episodes and it is 
difficult to identify such an event (but \citealt{Blanchard17}). 
Third, the AGN lifetime is short compared with the cosmic time, and
thus the contribution of those TDSs to the MBH mass growth and spin
evolution during the gas-cloud (AGN) accretion episode(s) is insignificant,
provided a more or less constant TDE rate for any individual galaxies.
Note also that we only consider the spin evolution of MBHs due to 
accretion of TDSs and gas-clouds, and ignore the effect of any other
processes and the intervals between any two adjacent accretion 
episodes. With such a setting, the MBH may be detected as a `TDE' 
when it accretes through the TDS channel, or as a normal `AGN',
possibly an NLS1, when it accretes via the standard thin disk at a
rate close to the Eddington limit through the gas-cloud accretion channel.


\subsection{Accretion of tidally disrupted stars}
\label{sec:TDE}

Normally a TDE occurs when a star moves so close to an MBH 
that the periapsis of its orbit $r_{\rm p}$ is less than the tidal radius
\begin{equation}
r_{\rm tid} = D_{*} \left( \frac{M_{\bullet}}{M_*}
\right)^{1/3}= 0.47 \ {\rm AU} \ x_* \left(
\frac{M_{\bullet,6}}{m_*} \right)^{1/3},
\end{equation}
where $D_*$ and $M_*$ are the radius and mass of the star,
respectively, $x_*$ is the radius of the star in unit of the solar
radius, $M_{\bullet,6}$ is the MBH mass in unit of $10^6\msun$, 
and $m_*$ is the stellar mass in unit of $\msun$. For simplicity, 
in this paper we assume that the TDSs all have the solar mass 
and the solar radius, i.e., $x_*=1$ and $m_*=1$. Assuming a 
mass spectrum for TDSs introduces little effect to our results 
presented in Section~\ref{sec:results}.

Once a star is tidally disrupted, roughly half of its mass is ejected, 
and the other half is bound to the MBH. The debris of the bound 
part starts to fall back to their pericenter after a time $t_{\rm min}\sim 
41 M_{\bullet,6}^{1/2} \textrm{ day}$ with a rate of
\begin{equation}
\dot{M}_{\rm fb} = \dot{M}_{\rm p} \left( \frac{t}{t_{\rm min}}
\right)^{-5/3},
\end{equation}
where 
$\dot{M}_{\rm p} =\msun /3t_{\rm min}\sim 1.9 \times 10^{26}
M_{\bullet,6}^{-1/2} \textrm{g~s}^{-1}$ is the peak of the fallback 
rate \citep[e.g.,][]{Lodato11}. The fallback rate is larger than the 
Eddington limit $\dot{M}_{\rm Edd} \simeq L_{\rm Edd}/\eta c^2 =
1.4\times 10^{24} M_{\bullet,6} (\eta/0.1)^{-1} \textrm{g~s}^{-1}$ at
the beginning if $M_{\bullet,6} \la 25$, where $\eta$ is the radiative 
efficiency of the MBH, and $L_{\rm Edd} = 1.3\times 10^{44} 
M_{\bullet,6} \textrm{erg~s}^{-1}$ is the Eddington luminosity. 
Therefore, the accretion of the debris of a TDS can be divided into 
two stages: an initial stage with $\dot{M}_{\rm fb} > \dot{M}_{\rm Edd}$,
in which the accretion may be via thick disk with outflows, and a second
stage with $\dot{M}_{\rm fb} < \dot{M}_{\rm Edd}$, in which the accretion 
may be via the standard thin disk.

It has been suggested that most of the debris may be expelled 
via outflows and the actual rate at the inner edge $r_{\rm in}$ of 
the disk is at most about the Eddington value, although the 
accretion rate exceeds the Eddington limit in the first stage 
\citep[e.g.,][]{Franchini16}. Therefore, we assume that the 
factor $f_\edd$ in the r.h.s. of Equation~(\ref{eq:dmdt}) is equal 
to $\min(\dot{M}_\fb/\dot{M}_\edd, 1)$ for the TDS accretion.

The disk formed by TDS debris may not be on the equatorial plane 
of the MBH initially since the stellar orbits of TDSs are probably 
isotropically distributed, as suggested by studies of loss cone stars 
surrounding massive binary black holes, in which those stars are 
found to be on chaotic orbits with random orientations \citep[see][]
{CY14, Vasiliev15}. On the one hand, the inner disk may also not 
be on the equatorial plane for this inclined TDS disk configuration 
(see Section~\ref{sec:tdespin}). The specific energy ($E_{\rm in}$) 
and angular momentum ($L_{\rm in}$) at the inner disk boundary 
(presumably the ISCO) have already been given in the literature 
for equatorial disks \citep[e.g.,][]{Bardeen72}.  However, those 
quantities for inclined disks are different from  the  equatorial case 
and are not explicitly given in the literature. Therefore, it is necessary 
to calculate $E_{\rm in}$ and $L_{\rm in}$  for disks with different 
inclination angles at the inner edge in order  to solve 
Equations~(\ref{eq:dmdt}) and (\ref{eq:spinevol}) that govern the 
MBH mass and spin evolution. On the other hand,  the misalignment 
between the angular momenta of the disk and MBH leads to the 
generation of torques on both the disk and MBH (the second term 
at the r.h.s. of Eq.~\ref{eq:spinevol}) and thus the precession and 
eventual alignment of them. Below we first introduce the procedures 
to calculate $r_{\rm in}$, $E_{\rm in}$, and $L_{\rm in}$ for an inclined 
disk in Section~\ref{subsec:EL}, then describe the precession and 
alignment of the TDS disk due to the disk angular momentum-MBH
spin coupling in Section~\ref{sec:tdespin}, and finally present the 
method to solve the mass and spin evolution equations in 
Section~\ref{subsec:method}.

\subsubsection{Specific energy and angular momentum at the inner boundary of an inclined disk}
\label{subsec:EL}

We adopt the Boyer-Lindquist coordinate system ($t$, $r$, $\theta'$, 
$\phi'$) to describe the MBH metric. For an MBH with any given spin 
$a$ and a non-equatorial disk with any inclination angle $i$,\footnote{
Here $i$ is defined as the angle between the orbital angular momentum 
of the disk at the inner edge and the MBH spin vector, and hence $i=0$, 
$\pi/2$, and $\pi$ correspond to the prograde equatorial orbit, polar orbit, 
and retrograde equatorial orbit, respectively.}
$r_{\rm in}$, $E_{\rm in}$ and $L_{\rm in}$ can be calculated by the 
following procedures. Note that similar method was adopted by 
\citet[][see also \citealt{Stone13}]{Hughes00, Hughes01} to calculate the
properties of non-equatorial circular orbits around Kerr black holes.

First, we calculate the ISCO (in unit of $r_{\rm g}$) for an equatorial disk  
according to \citep{Bardeen72}
\be
r_\isco &=&  3+Z_2 \mp[(3-Z_1)(3+Z_1+2Z_2)]^{\frac{1}{2}},  
\label{eq-risco}
\ee
where $Z_1 = 1+(1-a^2)^{1/3}[(1+a)^{1/3}+(1-a)^{1/3}]$,
$Z_2 = (3a^2+Z_1^2)^{1/2}$, and the upper/lower case of the 
``$\mp$'' (or ``$\pm$'') signs represents for prograde/retrograde 
orbits (the same afterwards).
The specific energy $E_{\rm ISCO}$ and angular momentum
$L_{\rm ISCO}$ at $r_{\rm ISCO}$ are then calculated through
\be
E(r) &=& \frac{r^{3/2}-2r^{1/2} + a}{r^{3/4}(r^{3/2}-3r^{1/2} +
2a)^{1/2}},  
\label{eq-er}\\
L(r) &=& \pm \frac{r^2 - 
2ar^{1/2}+a^2}{r^{3/4}(r^{3/2}-3r^{\frac{1}{2}} +
2a)^{\frac{1}{2}}},
\label{eq-phir}
\ee
by setting $r=r_\isco$.

Second, fix $r=r_{\rm ISCO}$ for prograde orbits, decrease the 
energy $E$ from the starting point $E_{\rm ISCO}$, and calculate 
the $z$-direction angular momentum $L_z$ and the Carter
constant $Q$ via 
\be
L_z(r,E) &=& {E (r^2 - a^2) - \Delta\sqrt{r^2(E^2 - 1) + r}\over
a(r - 1)}\;,
\label{eq-erl}
\ee
\be
Q(r,E) &=& {\left[(a^2 + r^2)E - |a|
L_z(r,E)\right]\over\Delta} - B\;,
\label{eq-qrl}
\ee
where $\Delta = r^2 - 2 r + a^2$ and $B\equiv r^2 + a^2 E^2 - 
2 |a| E L_z(r,E) +L_z(r,E)^2$ \citep[see][]{Hughes00, Hughes01}. 
We further check whether $d^2 R/d r^2$ equals zero; if not, then 
increase $r$ to repeat the above procedures until reaching $d^2 
R/d r^2 = 0$, the condition for the innermost circular orbit, or until 
an unphysical $L_z$ or $Q$ is reached. Here $R$ is defined as 
$R \equiv \left[E(r^2+a^2) -|a| L_z\right]^2- \Delta\left[r^2 + (L_z - 
|a| E)^2 + Q\right]$, and the orbital motion equations are $\Sigma^2 
(\d r/\d \tau)^2  = R$, $\Sigma^2 (\d \theta'/\d \tau)^2=Q-\cot^2\theta' 
L^2_z-a^2\cos^2\theta' (1-E^2)$, and $\Sigma(\d \phi'/\d \tau)=\csc^2
\theta' L_z+|a| E[(r^2+a^2)/\Delta-1]-a^2L_z/\Delta$,  with $\Sigma
\equiv r^2 +a^2\cos^2\theta'$. 
If $d^2 R/dr^2  =0$, then we record $r$, $E$, $Q$, $L_z$, and the 
inclination angle $i$. 

The total angular momentum for a non-equatorial plane circular orbit 
is not conserved because of the LT and frame-dragging effects. 
For example, the LT effect causes the orbit to precess around, i.e., 
$\theta'$ is not a constant. For non-equatorial circular orbits, it can be 
described as $L^2_{\rm tot}(\theta')=Q+L^2_z - a^2\cos^2 \theta' (1-E^2)$
\citep[e.g.,][]{defelice80, Frolov98}. Given a non-equatorial circular orbit 
with $r_{\rm in}$, $E$, $L_z$, and $Q$, we estimate the maximum and 
minimum values of $\theta'$, i.e., $\theta'_{\rm max}$ and $\theta'_{\rm 
min}$, by solving the motion equation $\Sigma d\theta'/ d\tau = 0$. Then 
we take $\left< L_{\rm tot} \right> =  [L_{\rm tot}(\theta'_{\rm max})
+2 L_{\rm tot}(\pi/2) + L_{\rm tot}(\theta'_{\rm min})]/4$ as the mean specific
angular momentum carried by the accreted material falling into the MBH
from a circular orbit with $r_{\rm in}$, $E_{\rm in}$, $L_z$, and $Q$, which
is also recorded for each non-equatorial circular orbit obtained above.

Third, decrease $E$ again, and repeat the above procedures until
$r=r_{\rm ISCO}$ for retrograde orbits. Hence an array of $r_{\rm
in}$, $E_{\rm in}$, $\left< L_{\rm tot} \right>$, $L_z$, and $i$ is obtained 
for the innermost equatorial and non-equatorial circular orbits around 
an MBH with any given spin $a$.

Finally, for a disk with any given inclination angle $i$, $r_{\rm in}$, $E_{\rm in}$
and $L_{\rm in}$ ($ \simeq \left< L_{\rm tot} \right>$) can be estimated 
by interpolation according to the array obtained above. Note here 
$\eta = 1-E_{\rm in}$.

\begin{figure} 
\centering
\includegraphics[width=3.4in]{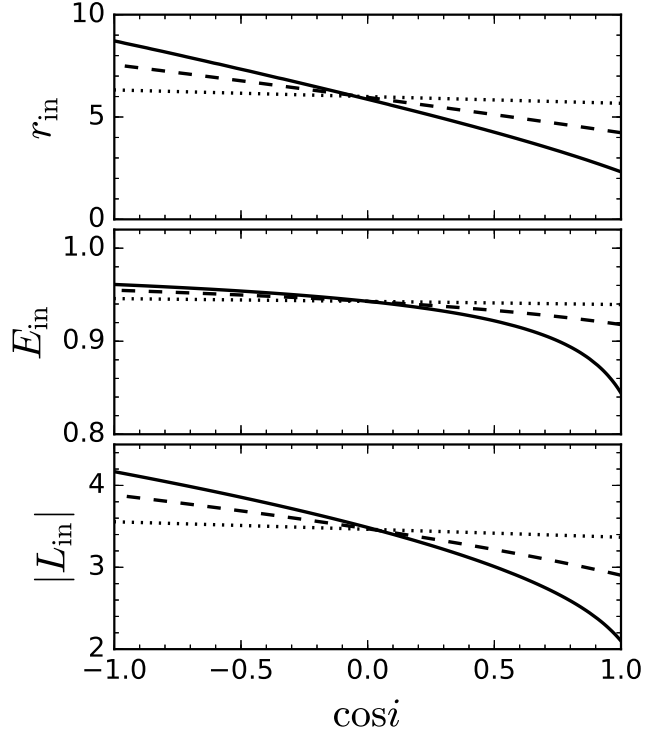}
\caption{The radius of the innermost circular orbit $r_\in$ (top panel), 
the specific  energy ($E_{\rm in}$; middle panel) and the specific angular 
momentum ($L_{\rm in}$; bottom panel) at $r_\in$, as a function of the
inclination angle $i$ for both equatorial and non-equatorial orbits. 
Solid, dashed and dotted lines represent for |a|=0.9, 0.5 and 0.1, respectively. 
According to our definitions, $\cos i=-1$ corresponds to retrograde equatorial 
disks, and $\cos i=1$ to prograde ones (see the footnote in Section~\ref{subsec:EL}).
}
\label{fig-le}
\end{figure}

Figure~\ref{fig-le} shows the radius of the ISCO $r_\in$, the specific energy 
and angular momentum at  $r_\in$ as a function of the inclination angle of 
the orbit for an MBH with $|a|=0.1$, $0.5$, and $0.9$, respectively. As seen 
from Figure~\ref{fig-le}, $r_\in$ decreases with increasing $\cos i$ and is in 
between that for retrograde ($i=180\arcdeg$) and prograde ($i=0\arcdeg$) 
orbits on the equatorial plane. The variation trends of $E_\in$ and $L_\in$ 
with the inclination angle are similar to that for $r_\in$.

\subsubsection{Disk angular momentum-MBH spin coupling: precession and alignment}
\label{sec:tdespin}

We introduce two Cartesian reference frames here, in order to study 
the precession and alignment of the disk. One is the observer's rest 
frame $Oxyz$ centered on the MBH, and the other is the rotating 
coordinate $Ox'y'z'$ centered on the MBH and the $z^\prime$-axis is 
always parallel with the MBH spin. Both of the Cartesian coordinates 
can be converted to spherical coordinates in terms of radius $r$, polar 
angle $\theta$ and azimuthal angle $\phi$. In this subsection, quantities 
(vectors and angles) with superscript `$'$' are in the $Ox'y'z'$ frame.
We assume that the initial direction of $\bf{J}_{\rm disk}$ for each TDS 
relative to the $Oxyz$ frame is randomly oriented, i.e., the distribution 
for the azimuthal angle $\phi$ is flat over $0$ to $2\pi$, and the 
distribution for the polar angle $\theta$ is proportional to $\sin\theta$
over $0$ to $\pi$.

The gravito-magnetic interaction between the MBH and an inclined 
TDS disk, at least at the initial super-Eddington accretion stage, will 
cause the precession and gradual alignment of their angular momenta, 
as detailed below.

{\bf (i) Global precession:} 
The 
angular momentum of the disk per unit area in the $Ox'y'z'$ reference 
frame can be expressed as 
\begin{equation}
{\bf L}' (r) =\Sigma \Omega r^2 r^2_{\rm g} \hat{\bm \lambda}',
\label{eq:ltot}
\end{equation}
where $\Sigma = 3\times10^6 \ \alpha^{-4/5}\ M^{1/5}_{\bh,6}\ \dot{m}^{3/5}\ 
(\eta/0.1)^{-3/5}\  r^{-3/5}\ (1-\sqrt{r_{\rm in}/r})^{3/5}\ \rm {g \ cm^{-2}}$ 
is the surface density of the disk with $\dot{m}=\dot{M}/\dot{M}_\edd$,  
$\Omega$ is the angular velocity, $\hat{\bm \lambda}'$ is a unit vector 
paralleled to ${\bf L}' (r)$ \citep{Franchini16}. The disk size is assumed 
to be twice the tidal radius, i.e., $r_{\rm out} = 2r_{\rm tid}/r_{\rm g}= 94 
M^{-2/3}_{\bullet,6}$.

Before $\dot{M}_\fb \leq \dot{M}_\edd$, the disk can still be warped, 
and the warp may propagate back and forth steadily inside the disk 
if $\alpha \lesssim H/R$ \citep[the bending wave regime; see][]{nelson99}. 
The behavior of the warp as a function of radius and time has been 
studied extensively, and it is quite well understood in the diffusive 
regime ($\alpha > H/R$) though there are still lots of work to do in 
the bending wave regime \citep[see][for a review]{Nixon16}. For the 
disk formed by a TDS, however, the warp could be small. It is therefore 
reasonable to assume that the disk angular momentum at different 
radii has the same direction at least in the initial super-Eddington 
stage \citep[see][]{Franchini16}, though further studies are needed 
to confirm this. In this case, the disk is assumed to precess globally 
as a rigid body  with a frequency \citep{Franchini16, sto12} 
\begin{equation}
\Omega_{\rm p} = \frac{\int^{r_{\rm out}}_{r_{\rm in}} \Omega_{\rm LT}
L(r) 2\pi r dr} {\int^{r_{\rm out}}_{r_{\rm in}} L(r) 2\pi rdr},
\label{eq:omegap}
\end{equation}
where the local LT precession frequency $\Omega_{\rm LT} = c^3 (4 a
r^{-3/2} - 3 a^2 r^{-2}) /2GM_\bullet (r^{3/2} +a)$. 
If a TDS disk orbits on a plane with normal direction $(\theta', \phi')$ in the $Ox'y'z'$ frame  
at time $t$, after a time-step $\delta t$, $\phi'$ becomes 
$\phi'(t+\delta t) = \phi' (t) + \Omega_{\rm p} \delta t$
due to global precession, while $\theta'$ keeps the same.

{\bf (ii) Alignment:}
An initial inclined disk will be gradually dragged to the equatorial plane 
of the MBH owing to the disk viscosity, leading to an alignment between 
the angular momenta of the disk and MBH. 
\citet{Franchini16} have shown that the alignment timescale $t_\al$
depends on the MBH mass and spin, and they find that their results 
are consistent with that obtained by the method provided in \citet{fou14}. 
\citet{Franchini16} showed that $t_\al \propto 10^{-3.7a}$ for $10^7 \msun$ 
MBHs in their Figure~12, and we further find that $t_\al \propto M^{0.35}_\bh$
according to Equation~(35) in \citet{fou14}. Combining the above two results, 
$t_{\rm al}$ can be approximated as
\begin{equation}
\log (t_{\rm al}/\textrm{day} ) \sim -3.7 a + 3.9+ 0.35 (\log M_{\bh, 6}-1),
\end{equation}
for a disk with a fixed viscosity parameter $\alpha =
0.1$.\footnote{According to \citet{Franchini16}, $t_{\rm al} \propto
\alpha^{-1}$. Therefore, we have also checked whether our results on
the spin distributions are affected by choosing a different $\alpha$
and find that choosing an $\alpha$ a few times smaller or
larger introduces insignificant effect to the resulting spin distributions.}
The alignment timescale can range from days to years depending 
on the MBH spin, and the higher the spin, the faster the alignment.
The viscous time [$t_{\rm vis} \sim \alpha^{-1} (GM_\bullet/R^3)^{-1/2} 
(H/R)^{-2}$] at the outer disk edge is on the order of years for the 
super-Eddington stage \citep{strubbe09}, which could be longer or 
comparable to the alignment timescale. The time for the accretion 
rate to decline to sub-Eddington value is about the same order, which 
means the disk could still be inclined in some cases when the disk 
becomes thin and enters the diffusive regime.

With the assumptions made above, the projected disk angular momentum 
per unit area in the $x'y'$-plane approximately evolves as 
\begin{equation}
L^\prime_{xy} = L^\prime_{xy,0} \exp (-t/t_{\rm al}),
\label{eq:Lxy}
\end{equation}
where $L^\prime_{xy,0}$ is the initial value \citep{Franchini16}.

\subsubsection{Numerical method to solve the MBH spin evolution equation}
\label{subsec:method}

In this sub-section, we introduce the numerical method to solve 
Equation~(\ref{eq:dmdt}), and Equation~(\ref{eq:spinevol}) or 
(\ref{eq:dadt}). As discussed above, the disk formed by each TDS 
is initially thick in geometry due to super-Eddington accretion. In 
this stage, the disk is assumed to precess globally and gradually 
align with the MBH equatorial plane. After that, the disk becomes 
thin when the accretion rate declines to $\dot{M}_\edd$, and there 
are two possibilities: (1) the disk angular momentum is already 
aligned with the MBH spin; (2) the disk is not aligned yet and will 
be warped under the diffusive regime. For the second case, if the 
warp radius exceeds the disk size, then the disk angular momentum 
is assumed to be simultaneously aligned with the MBH spin. 
Following \citet{per09}, the rapid temporal evolution of the warped 
disk is separated from the longer temporal evolution of the MBH. 
In particular, the disk is assumed to transit through a sequence 
of steady states over a period of $n \delta t < t < (n+1) \delta t$
 before the alignment of disk and MBH angular momenta (adiabatic 
 approximation). Here $ \delta t \ll t_{\rm al}$, and  $n$ is an integer 
 of $0, 1, 2, ...$. Within each time-step  $\delta t$, the inclination 
 angle $i$ is approximately  unchanged, and $E_{\rm in}$ and 
 $L_{\rm in}$ can be  estimated according to the array obtained in 
 Section~\ref{subsec:EL}.

The spin direction of an MBH is described by $\theta_\bullet$ and $\phi_\bullet$ 
with respect to the observer in the $Oxyz$ frame. Then the MBH angular 
momentum can be expressed as ${\bf J}=(J_x,\ J_y,\ J_z)=J (\sin\theta_\bullet 
\cos \phi_\bullet,\ \sin \theta_\bullet\sin\phi_\bullet,\ \cos\theta_\bullet)$, and 
${\bf J'}$ can be related to ${\bf J}$ by a rotation matrix 
$\mathcal{R}_{ij}$ ($i,\ j$=1, 2, 3), i.e., ${\bf J}^\prime={\bf J} \mathcal{R} =[0, 0, J]$. 
Similarly, for a TDS orbiting with angular momentum direction $\theta$ and $\phi$ 
relative to the observer, the angular momentum per unit area is described by ${\bf L}
= L(\sin\theta\cos\phi,\ \sin\theta\sin\phi,\ \cos\theta)$, and ${\bf L}^\prime=L \hat{\bf l}' 
= {\bf L} \mathcal{R} $, from which the unit vector $\hat{\bf l}'=(\hat{l}'_x,\ \hat{l}'_y,\ 
\hat{l}'_z)$ can be derived. This corresponds to an inclination angle $i=\theta'=\arccos
(\hat{l}'_z)$ between the disk and MBH angular momenta, and an azimuthal angle 
$\phi'=\arccos \left(\hat{l}'_x/\sqrt{\hat{l}'^2_x+\hat{l}'^2_y} \right)$ in the $Ox'y'z'$ 
reference frame.

Considering Equation~(\ref{eq:spinevol}) in the $Ox'y'z'$ frame, within 
$\delta t$, the spin change in terms of three components $\delta {\bf J}'
=(\delta J'_x,\ \delta J'_y,\ \delta J'_z)$ can be calculated through 
\be
\begin{split}
\delta J'_x & \approx   \delta t  \left[  \dot{M} r_{\rm g} c
L_\in \hat{l}'_x + \frac{4\pi G J}{c^2} \int_{\disk} \frac{
L'_y(r)}{r^2 r_{\rm g}} \d r \right],  \\ 
\delta J'_y & \approx  \delta t  \left[  \dot{M} r_{\rm g} c L_\in
\hat{l}'_y - \frac{4\pi G J}{c^2} \int_{\disk}
\frac{L'_x(r)}{r^2r_{\rm g} } \d r \right],   \\ 
\delta J'_z & \approx  \delta t  \left[  \dot{M} r_{\rm g} c L_\in
\hat{l}'_z \right].  
\end{split}
\label{eq:jxyz}
\ee
The spin variation in the observer's $Oxyz$ frame can be obtained via 
$\delta {\bf J}=\delta {\bf J}' \mathcal{R}^{-1}$. In the next time-step, $\phi'$ 
of the disk angular momentum is updated due to the global precession, 
and the inclination angle $i$ is updated owing to the exponential decline 
of $L'_{xy}$ with time (see Section~\ref{sec:tdespin}), i.e., 
\begin{equation}
i = \arccos \left(\sqrt{1-L'^2_{xy}/L'^2}\right).
\end{equation} 
This leads to an update of $\hat{\bf{l}}'=(\sin i \cos\phi'$, $\sin i \sin\phi'$, 
$\cos i)$, and Equation~(\ref{eq:jxyz}) is solved once again. The above 
procedures are repeated until $|\cos i'|$ is close to 1 or the accretion rate 
falls below the Eddington limit. 

If the disk angular momentum is still not aligned with the MBH spin when 
$\dot{M}_{\rm fb} < \dot{M}_{\rm Edd}$, then the disk enters the diffusive 
regime and warps.  The disk is maximally warped at around the warp radius 
$r_\wp$ \citep[see][for detailed expression of $r_\wp$ ]{per09}.
If the disk size is larger than $r_\wp$, then the spin evolution is 
still described by Equation~(\ref{eq:spinevol}), where $\hat{\bf{l}}$ 
is now parallel with the MBH spin. The analytical solution for the disk 
profile that describes how the disk is warped at different radii is given 
in \citet{mar07} relative to the MBH coordinate, and the analytical 
expression of the torque term in Equation~(\ref{eq:spinevol}) is 
directly given by \citet[][see their Appendix]{per09} in the $Ox'y'z'$ 
frame, which is then rotated back to the $Oxyz$ frame. 
However, if the warp radius exceeds the disk size, i.e., $r_\wp>r_{\rm out}$, 
which means there is no warp in the disk, then the MBH and the disk are 
assumed to be instantaneously aligned (or anti-aligned) with each other 
\citep[for the criterion for anti-alignment, see Section~\ref{sec:spinevol}; 
see also][]{kin05}. For both cases, the second term on the r.h.s. of 
Equation\,(\ref{eq:spinevol}) reduces to 0. Then we only need to solve 
Equation~(\ref{eq:dadt}) governing the spin magnitude evolution, and the 
spin direction is realigned to that of ${\bf J}_{\rm disk}+{\bf J}_\bullet$. 
For the TDS accretion, $r_\wp > r_{\rm out}$ is frequently satisfied when 
$\dot{M}_{\rm fb} < \dot{M}_{\rm Edd}$.

\begin{figure} 
	\centering
	\includegraphics[width=3.4in]{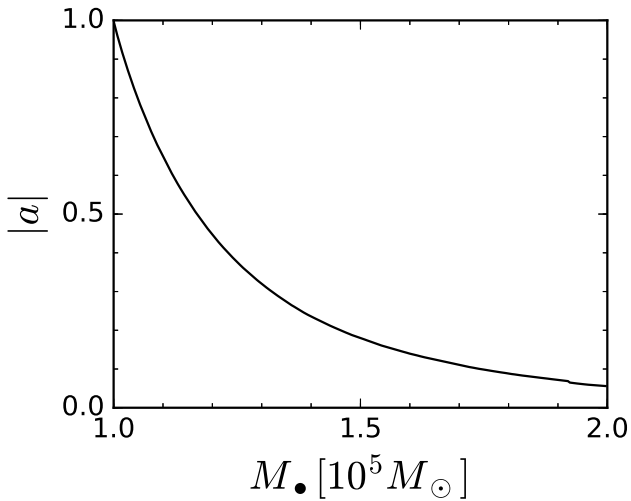}
	\caption{Spin evolution due to consecutive accretion of TDSs for an MBH with
	initial conditions of $M_{\bh, 0}=10^5 \msun$ and $a_0=0.998$. 
	The MBH mass is used as a surrogate of accretion time.}
	\label{fig-am_tde}
\end{figure}

Figure~\ref{fig-am_tde} shows the spin evolution of an MBH with 
its mass increase due to consecutive accretion of TDSs. Here we 
only show the evolution of the absolute spin value $\left| a \right|$, 
though the spin direction also evolves with time. The orbital 
orientations of the injected TDSs are assumed to be randomly 
distributed, which can result in almost the same probability for  initial
prograde and retrograde accretion.\footnote{Note that the cross section 
for the tidal disruption of retrograde stars around rapidly rotating MBHs 
can be somewhat larger than that of prograde stars \citep{Beloborodov92}, 
which may lead to even more significant spin-down effect. } About half of 
the TDSs should be accreted onto the central MBH via prograde direction 
and the MBH spin increases during each of those accretion events. The 
other half are accreted onto the MBH via retrograde direction and the  
MBH spin decreases in this case. The injection of negative angular 
momentum is more efficient than that of the positive one because the inner 
disk radius for a retrograde orbit is larger than that for a prograde orbit (see 
Fig.~\ref{fig-le}). Therefore, the MBH spin appears to be quickly spun down 
to $\sim0$ when the MBH mass increases by a factor of $\sim 2$ due to 
consecutive accretion of TDSs. During the period of a single prograde 
TDS accretion, the MBH spin does increase, but the total amount of the
accreted mass in a single prograde TDE is too small and thus the 
spin increase is too slight to be clearly shown in Figure~\ref{fig-am_tde}.

\subsection{Chaotic accretion via thin disk}
\label{sec:subsecchao}

AGN-like nuclear activities may contribute significantly to the 
growth of MBHs with mass $\sim 10^6\msun$, as many AGNs 
are detected with central MBH mass around this value. Therefore, 
we also consider their effects on the spin evolution of the central MBHs. 

Observations of maser disks, molecular clouds, and Galactic 
star clusters may provide clues about the gas-cloud mass in 
individual AGN accretion episode. The sub-pc maser disk in 
NGC 4258, with a $\sim 4 \times 10^7\msun$ MBH 
\citep{menezes18}, is estimated to have a mass in the range 
from $\sim 10^4 \msun$ to $\sim 10^6 \msun$ \citep[e.g.,][]{her98, 
her05}. IC 2560 hosts a $\sim 3\times10^6 \msun$ MBH and 
a maser disk of size $\sim 0.2$\,pc \citep{yamauchi12}. According 
to the estimated accretion rate of IC 2560 ($\sim10^{-5} \msun\,{\rm 
yr}^{-1}$; \citealt{ishihara02}), the disk mass is about $10^4 \msun$ 
assuming the standard thin disk model \citep{sha73}. For the 
$4\times 10^6\msun$ MBH in our own Galactic center, observations 
suggest that it was active six million years ago and the mass of the 
disk in that accretion episode is $\ga 10^4\msun$ 
\citep[e.g.,][]{Nayakshin05, Levin03, Paumard06, LuJ09}. 
In addition, star clusters in the Galactic center such as Arches 
and Quintuplet are found to have mass of $\sim10^4\msun$ 
\citep[e.g.,][]{figer99, habibi13}. Dynamical simulations also 
suggests that the underlying stellar mass of Arches is well within 
$10^4 \msun$ and $10^5 \msun$ \citep[e.g.,][]{harfst10}. These 
findings indicate that the molecular clouds which later collapse 
and form those star clusters probably have the masses $\gtrsim 
10^4 \msun$. 
Moreover, observations found that the molecular clouds in our 
Milky Way typically have masses $\sim 10^4$-$10^6 \msun$ 
\citep[for a review see][]{heyer15}.

The above lines of observational constraints on gas-cloud and 
disk masses may suggest that the gas-cloud mass in each AGN 
accretion episode is in the range of $10^4$-$10^6\msun$. One 
may note that the accretion disk may not be too massive, otherwise 
it may fragment into clumps against its own gravity \citep[][]{kol80, 
Goodman03, king07}. The self-gravitating disk mass $M_{\rm sg}$ 
can be estimated via the Toomre's $Q$ criterion, i.e., the disk size 
is limited by the fragmentation radius where $Q\sim c_s\Omega/\pi 
G\Sigma=1$, with $c_s$ the sound speed, $\Omega$  the Keplerian 
angular velocity, and $\Sigma$ the disk surface density. This results 
in an upper limit on the disk mass \citep[][]{Dotti13},
i.e.,
\be
M_{\rm sg} \approx 2\times10^4 \alpha^{-1/45}_{0.1} \left(\frac{f_\edd}{\eta_{0.1}} 
\right)^{4/45} M^{34/45}_{\bullet,6} \msun,
\ee
where $\alpha_{0.1}=\alpha/0.1$ and $\eta_{0.1}=\eta/0.1$.

If the cloud mass is about $10^4 -10^5 \msun$, either or not regulated 
by self-gravity, then the growth history of an MBH with final mass $\ga 
10^{6}\msun$ should have a number of or many AGN accretion episodes
if the contribution of these phases to the mass growth is significant, and 
the accretion in different AGN episodes is probably chaotic, i.e., with 
randomly distributed inclination angles relative to the MBH spin. We 
therefore consider the following cases for disk mass in the AGN episodes.

(i) The whole gas-cloud collapse and form an accretion disk without 
mass loss, i.e., $M_\disk=M_\cl$, where $M_\cl$ is set to be constant 
$10^4 \msun$ or $10^5 \msun$ for all MBHs. 

(ii) The disk mass is regulated by self-gravity, i.e., $M_\disk=\min(
M_\cl,\ M_{\rm sg})$, where $M_\cl$ is $10^4 \msun$ or $10^5 \msun$ 
for all MBHs, or randomly drawn from a flat distribution between $10^4 
\msun$ and $10^5 \msun$ in the logarithmic space.

For both cases, the gas-cloud in each episode is assumed to fall 
in with random directions, and similar to the disk formed by TDSs, 
initially $\bf{J}_\disk$ has an azimuthal angle $\phi$ randomly 
distributed, and a polar angle $\theta$ drawn from a distribution 
proportional to $\sin\theta$. The disk is assumed to be described 
by the standard thin disk model, and the radial and vertical shear 
viscosities follow a power law, i.e., $\nu_1 \propto r^{3/4}$ and 
$\nu_2 \propto r^{3/4}$. For radial viscosity $\nu_1$, the $\alpha$ 
prescription is adopted with $\alpha=0.09$, and the main properties 
of the disk are detailed in \citet{per09} and \citet{Dotti13}.

Similar to the late stage of sub-Eddington accretion of TDSs 
described in Section~\ref{sec:TDE}, the mass and spin evolution 
of MBHs are obtained by solving Equation~(\ref{eq:dmdt}) and 
Equation~(\ref{eq:spinevol}) or (\ref{eq:dadt}) with adiabatic 
approximation \citep{per09}. The Eddington ratio $f_\edd$ in 
Equation~(\ref{eq:dmdt}) mainly determines the mass growth 
rate and has little effect on the spin-mass evolutionary curves. 
We therefore assume a constant $f_\edd$ of $0.3$ for the 
gas-cloud accretion. As the inner disk is bent to the MBH 
equatorial plane due to the Bardeen-Petterson effect, 
$\hat{\mathbf{l}}$ in Equation~(\ref{eq:spinevol}) is parallel to 
the MBH spin, and the ISCO is the value for equatorial thin disks.

\begin{figure} 
	\centering
	\includegraphics[width=3.4in]{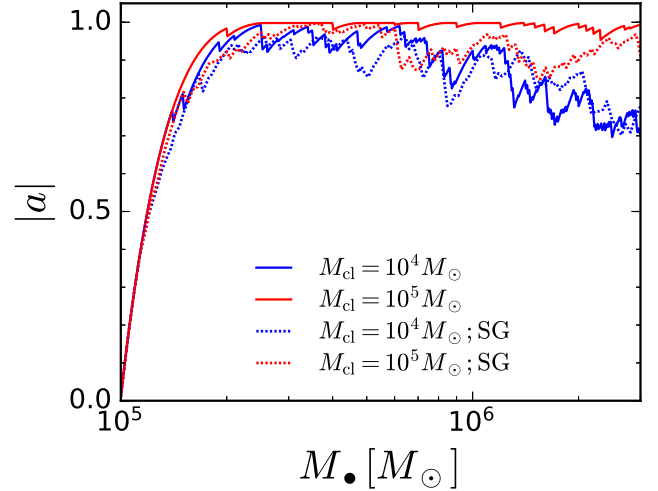}
	\caption{Spin evolution due to chaotic accretion of gas-cloud 
	for initially non-spinning MBHs with $M_{\bh,0}=10^5\msun$. 
	The red and blue solid (or dotted) lines are for a gas-cloud 
	mass of $10^5 \msun$ and $10^4 \msun$ in each gas-cloud 
	accretion 	episode without (or with) consideration of disk self-gravity.
}
	\label{fig-am_cl}
\end{figure}

Figure~\ref{fig-am_cl} shows the spin evolution due to chaotic accretion 
of gas clouds. The cloud mass is set to be constant $10^4 \msun$ or   
$10^5 \msun$ across the accretion history, and the disk mass is either 
or not capped by the self-gravitating mass.For an initially non-rotating 
MBH of $10^5 \msun$, all of the cases show an initial sharp increase 
in spin, which is due to quick alignment of MBH spin to the disk angular 
momentum (even if the disk is initially counter-rotating), because the disk 
angular momentum dominates over the MBH spin at the beginning 
\citep[see also][]{Dotti13}. With the growth of MBHs, $J_\bh$ becomes 
comparable with $J_{\rm disk}$ and the alignment time becomes 
significant with respect to the accretion time. This results in a decrease 
in spin more or less due to the more effective injection of negative 
angular momentum by counter-rotating disks. The spin-down effect 
is more prominent for low-mass clouds ($10^4 \msun$), because 
the alignment of MBH spin to disk angular momentum is less efficient. 
If we compare the alignment timescale $t_{\rm al}$ \citep{per09} with 
the disk-consumption time $t_{\rm acc}$ in each episode, we find that 
$t_{\rm acc}/t_{\rm al} \sim 5 M_{\cl, 4} M^{-33/35}_{\bh, 6}$ by 
neglecting some unimportant terms. For MBHs of $\sim 10^5 \msun$, 
$t_{\rm acc}/t_{\rm al} \gg 1$, and the alignment is efficient even for 
low-mass clouds ($10^4 \msun$). The disk mass and size are crucial 
to the second term of Equation~(\ref{eq:spinevol}) and determines how 
efficient the alignment is. The larger the disk mass is, the higher spin 
the MBH maintains.
Considering the self-gravity of disks does not make much difference 
to the spin evolution for the case with $M_\cl=10^4 \msun$ though it 
does introduce some effect to the case with $M_\cl=10^5\msun$ 
because $M_{\rm sg}$ is substantially smaller than $M_\cl$ only when 
$M_\cl$ exceeds $10^5\msun$ at $M_\bullet\sim 10^6\msun$ 
(Fig.~\ref{fig-am_cl}, and see also Fig.~11 in \citealt{zhanglu19}).

\section{Model settings and model results}
\label{sec:results}

We consider MBHs all with assumed initial masses of $M_{\bh, 0}=10^5 
\msun$, initial spins of $a_0=0.5$,\footnote{We have checked and found 
that choosing a different initial spin distribution does not significantly affect 
the results presented in this paper.} and final masses of $M_{\bh,\f} =10^{6.5} 
\msun$. They experience accretion of both gas-clouds and TDSs. During the 
period of gas-cloud accretion, they appear as AGNs. Our model involves two 
parameters, i.e., {\boldmath $f_\tde$} and {\boldmath $M_\cl$}, as detailed below.

\noindent
(1) The parameter $f_\tde$ denotes the fraction of mass growth 
of the final MBH contributed by TDSs, i.e., $f_\tde \equiv \Delta 
M_{\bh,\tde}/(M_{\bh,\f}-M_{\bh,0})$ with $\Delta M_{\bh,\tde}$ 
the total mass increase contributed by TDS accretion. In reality 
$f_\tde$ may not be a constant but depend on the MBH mass 
and/or the environment. However, how $f_\tde$ explicitly 
depends on MBH mass is currently unknown. As the main focus 
of this paper is to demonstrate the effect of accreting TDSs on 
the spin evolution of low mass MBHs ($\sim 10^6 \msun$), it is 
reasonable to assume a constant $f_\tde$, at least as the mean 
value. Therefore, we consider three cases with fixed values of 
$f_\tde$ for all MBHs, i.e., $f_\tde=0.1$, $0.5$, and $0.9$, 
respectively. These three cases correspond to that the gas-cloud 
accretion dominates the MBH growth, the accretion of gas-cloud 
and TDSs are equally important, and the TDS accretion dominates 
the MBH growth, respectively. It is also plausible that $f_\tde$ is 
different for MBHs with the same mass due to their different 
environments. For this reason, we also consider a case with $f_\tde$ 
randomly distributed between $0.1$ and $0.9$ for those MBHs. 

\noindent
(2) The parameter $M_\cl$ denotes the gas-cloud mass available 
in each gas-cloud (AGN) accretion episode. Our models assume 
that $M_\cl$ is either fixed at $10^4 \msun$ or $10^5 \msun$ for 
each gas-cloud accretion episodes of all the MBHs, or it is randomly
distributed over $10^4 \msun$ and $10^5 \msun$ for different episodes 
of each MBH. For those fixed $M_\cl$ models, the disk is either or not 
regulated by self-gravity (for these settings, see discussions on 
observational results in Section~\ref{sec:subsecchao}).

For an MBH with fixed $M_\cl$ during its accretion history, the 
number of gas-cloud accretion episodes is calculated through 
$N_\cl={\rm int} [(1-f_\tde)(M_{\bh, \f}-M_{\bh,0})/M_\cl]+1$. 
We assume that each MBH starts with accreting TDSs and we 
ignore the intervals between any two adjacent accretion episodes. 
The gas-cloud accretion episodes are randomly injected into the 
growth history of an MBH. Specifically, we produce $N_\cl$ random 
numbers between $M_{\bullet,0}$ and $M_{\bullet,\rm f}$. With the 
growth of the MBH due to TDS accretion, once the mass reaches 
any one of those random values, it starts a gas-cloud accretion episode 
and the TDS accretion is switched off. This gas-cloud accretion 
episode ends when the gas-cloud is all consumed, and the TDS 
accretion is then switched on again until the MBH mass reached 
another random number produced above and another gas-cloud 
(AGN) accretion episode is switched on. 

For those cases considering disk self-gravity, the number of gas-cloud 
accretion is different even for accretion histories with the same $f_\tde$ 
since $M_\sg$ is dependent on the MBH mass. In the modeling, we 
randomly produce $1$, $2$, $3$, ... random numbers between 
$M_{\bullet,0}$ and $M_{\bullet,\rm f}$, calculate $M_{\rm sg}$ and 
$M_\disk=\min(M_{\rm sg}, M_\cl)$ for each random number produced, 
and get the sum of those disk masses. The above procedures are 
terminated once the summation of the disk masses exceeds 
$(1-f_\tde)(M_{\bh, \f}-M_{\bh,0})$. Then the accretion history of the 
MBH is known, i.e., when the MBH starts a period of gas-cloud accretion, 
and how massive the disk mass is in each episode.  
Therefore, different models have different 
numbers of gas-cloud accretion episodes. For a model even with a 
fixed $N_\cl$, the accretion histories of individual MBHs are 
still different from each other since the TDS and gas-cloud accretion episodes
are injected into the accretion history randomly and each with a random orientation (see 
Section~\ref{sec:spinevol}). 

According to our settings, we have $20$ models in total (see Table~\ref{tbl-2}). 
For each model, we calculate the mass and spin evolution for $500$ MBHs 
according to the methodology described in Section~\ref{sec:spinevol}. From 
each of the evolutionary curve, we randomly select 1000 mock AGNs 
with $M_\bh \sim 10^{5.5} - 10^{6.5}\msun$, and we also randomly select 
1000 mock TDEs with the same mass range. According to these procedures,  
the differential spin distribution of the MBHs in both the mock AGNs and TDEs 
can be obtained, i.e., $dn/da=N^{-1}_{\rm tot} \Delta N/ \Delta a$. Here 
$N_{\rm tot}= 500,000$ is the total number of objects included in the mock sample(s), 
and $\Delta N$ is the number of MBHs with spins between $a$ and $a + \Delta a$.

\begin{figure} 
\centering
\includegraphics[width=3.4in]{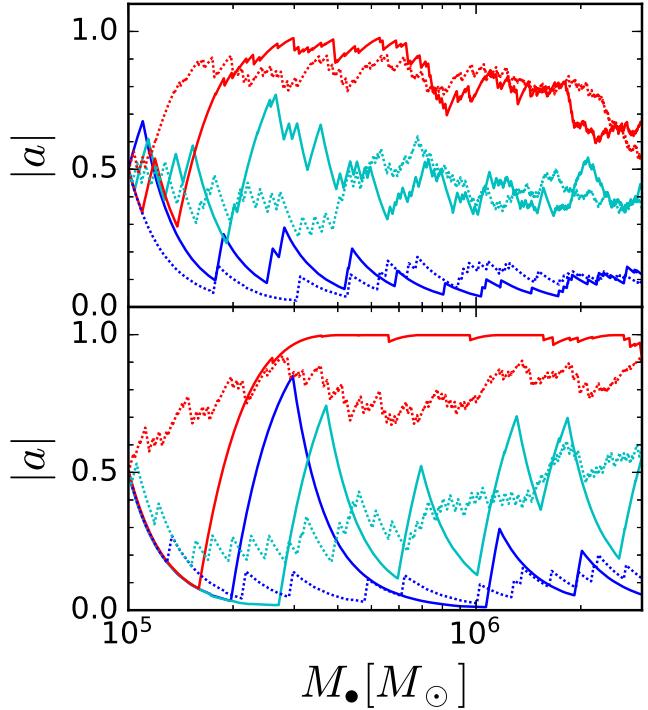}
\caption{Examples of spin evolutionary curves. The gas-cloud in each 
AGN accretion episode is $10^4 \msun$ ($10^5\msun$) in the top 
(bottom) panel, and the solid (or dotted) lines in each panel are for disks 
without (or with) consideration of self-gravity. The fraction of MBH mass 
growth contributed by TDSs is set as $10\%$ (red), $50\%$ (cyan), or 
$90\%$ (blue).}
\label{fig-am}
\end{figure}

Figure~\ref{fig-am} shows some spin evolutionary curves, as examples, 
for MBHs experiencing both TDS and AGN accretion episodes, with 
different $f_\tde$ and $M_\cl$, and with or without consideration of 
self-gravity in the gas-cloud accretion phase. As shown in this figure, 
those MBHs are spun up in the episodes of gas-cloud accretion and 
the MBH spins can increase to large values in a single gas-cloud 
accretion episode when the disk is massive (e.g., $10^5 \msun$) 
and the disk self-gravity is ignored, while the MBHs are spun down 
significantly after the accretion of a large number of TDSs.
This is because each TDS adds at most $\sim 0.5 \msun$ to the MBH 
mass, while a gas-cloud accretion episode adds $\sim10^4$-$10^5\msun$. 
Therefore, the angular momentum of the TDS is tiny compared to the MBH 
spin, and the probability of co- or counter-alignment to the MBH spin is essentially 
the same for the disk angular momentum. Gas-clouds, however, have much 
higher (orbital) angular momenta, which can dominate over that of the MBH, 
and therefore the MBH spin tends to co-align with the disk angular momentum.
For a larger $M_\cl$ ($>10^5\msun$), the alignment of the MBH spin 
to the disk angular momentum is more efficient, and the spin can reach 
even higher values in a single chaotic accretion episode. How the MBH 
spin evolves depends on the relative contribution from accretion of TDSs 
to the MBH mass growth. If $f_\tde= 0.1$, the MBH spin can maintain at 
high values at later stages when $M_\bh \ga 10^6\msun$; if $f_\tde = 0.9$, 
the MBH spin may be always close to $0$ when $M_\bh \ga 10^6\msun$. 
However, if considering the disk self-gravity, the probability for MBHs to 
be spun up to large values may be significantly suppressed in a single 
gas-cloud accretion episode, especially for those cases with $M_\cl \gtrsim 
10^5\msun$, because the disk mass is limited to about $M_\sg \sim 10^{4}
\msun$ at $M_\bh \sim 10^6\msun$. This would lead to a lack of mock MBHs 
with spins close to the maximum spin value as will be shown in Figure~\ref{fig-pa_sg}.

\begin{table}
\begin{center}
\caption{Spin measurements of individual AGNs via Fe K$\alpha$ lines.}
	\label{tbl-1}
	\begin{tabular}{ccc} 
    \tableline
    Object name     & $M_\bullet (10^6 \msun)$    &  Spin \\
    \tableline
    1H 0707-495     & $\sim2.3$                     & $>0.97$ \\
    Ark 564         & $\sim1.1$                     & $0.96^{+0.01}_{-0.07}$ \\
    MCG 6-30-15     & $2.9^{+1.8}_{-1.6}$           & $>0.98$ \\
    Mrk 359         & $\sim1.1$                     & $0.66^{+0.30}_{-0.54}$ \\
    NGC 1365        & $\sim2$                       & $\ge0.84$ \\
    NGC 4051        & $1.91\pm0.78$                 & $>0.99$ \\
    \tableline
	\end{tabular}
	\end{center}
	\tablecomments{This table lists those individual AGNs with spin 
		measurements that have MBH mass in the range of $10^6-10^{6.5}\msun$, 
		selected from \citet{Brenneman13} and \citet{Reynolds14}. 
		References for those spin measurements are \citet{zoghbi10} 
		for 1H 0707-495, \citet{walton13} for Ark 564 and Mrk 359, \citet{BR06} 
		for MCG 6-30-15, \citet{risaliti13} for NGC 1365, and \citet{patrick12} 
		for NGC 4051, respectively. The mass measurements for them are adopted from
		\citet{ZW05} (1H 0707-495, Ark 564 and Mrk 359),  \citet{mchardy05} (MCG 6-30-15), 
		\citet{risaliti09} (NGC 1365), and \citet{peterson04} (NGC 4051). }
\end{table}

\begin{figure*} 
\centering
\includegraphics[scale=0.85]{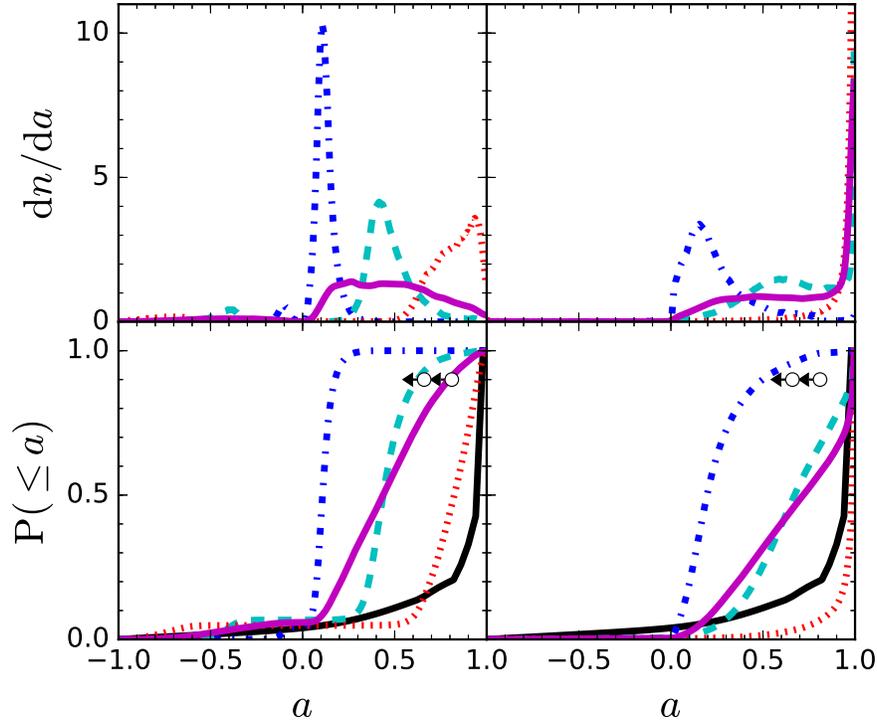}
\caption{Differential (upper panels) and cumulative (lower panels) spin 
distributions of those mock MBHs ``detected'' in the gas-cloud (AGN)
accretion episodes obtained from simulations for different models. Left 
(right) panels represent models with fixed cloud mass of $10^4 \msun$ 
($10^5 \msun$) in each AGN episode. The fraction of MBH mass 
growth due to TDS accretion is set as $10\%$ (red dotted), $50\%$ 
(cyan dashed), $90\%$ (blue dot-dashed) or randomly distributed 
between $10\%$ to $90\%$ over different MBHs (solid magenta). 
The black line shows the cumulative spin distribution inferred from 
spin measurements listed in Table~\ref{tbl-1}, where the errors have 
been taken into account. The open circles represent the constraints 
on the effective spin obtained from the stacked X-ray spectrum of 
$51$ NLS1s, i.e., $a<0.81$ or $<0.66$ at the $90\%$ confidence 
level, presented in \citet{Liuetal15}.
}
\label{fig-pa}
\end{figure*}

\begin{figure*} 
\centering
\includegraphics[scale=0.85]{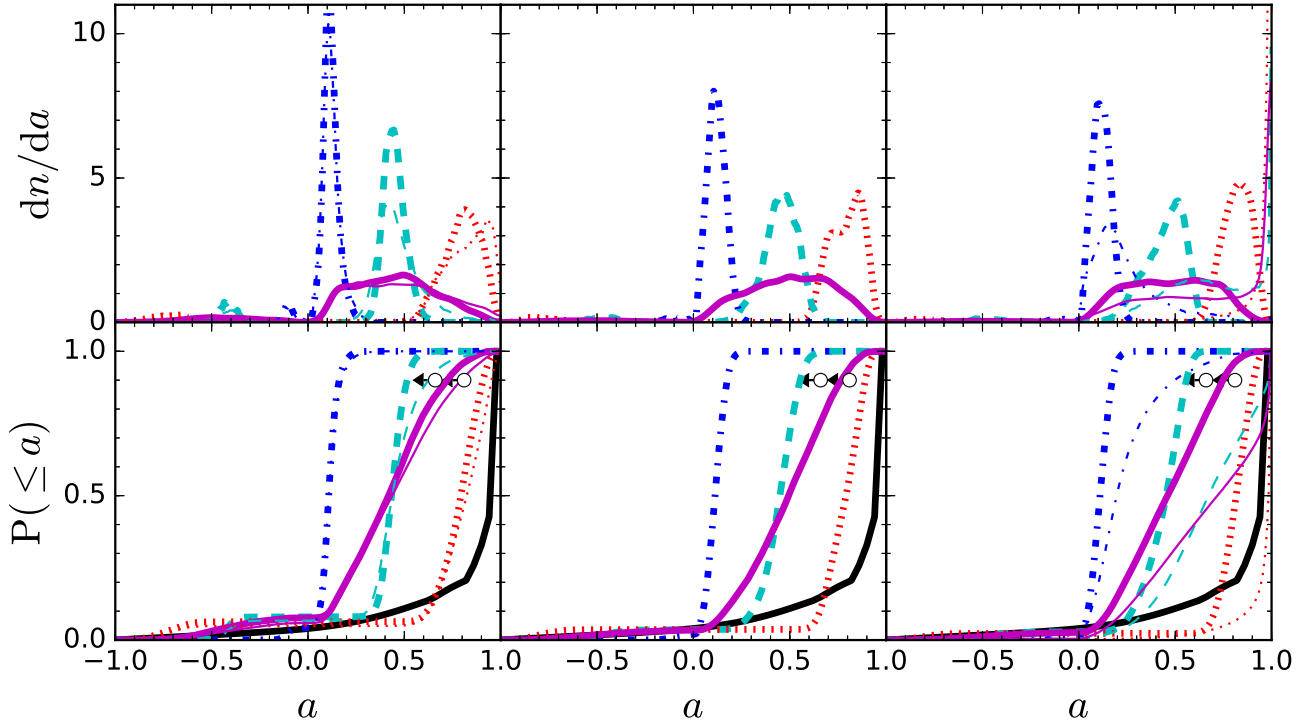}
\caption{Legend similar to Figure~\ref{fig-pa}. Thick curves in each
panel represent the results obtained from those models by considering
the self-gravity regulation of the disks in the gas-cloud accretion
episodes, i.e., $M_\disk=\min(M_{\rm sg},\ M_\cl)$ (see
Table~\ref{tbl-2}). Left and right panels show the cases with
assumption of 
$M_\cl=10^4 \msun$ and $10^5 \msun$, respectively; middle panels show
the cases with assumption of $M_\cl$ randomly distributed over $10^4 \msun$ and
$10^5 \msun$. For comparison, the results shown in Fig.~\ref{fig-pa}
for models without considering the self-gravity regulation of the
disks are also correspondingly shown here by thin curves in the left
and right panels.}
\label{fig-pa_sg}
\end{figure*}

Figure~\ref{fig-pa} shows differential (top panels) and cumulative 
(bottom panels) distributions of spins for those mock samples of 
MBHs selected in the gas-cloud (AGN) accretion episodes, 
according to the mass and spin evolutionary curves obtained 
from our simulations by setting different $f_\tde$ and $M_\cl$ 
without consideration of disk self-gravity. Here we show both the 
prograde-spin accretion ($a>0$) and retrograde one ($a<0$). It 
appears that retrograde-spin accretion is almost negligible for all 
models simply because the alignment timescale is short and 
anti-alignment cases rarely occur. If $M_\cl=10^4 \msun$ (left 
panels), the MBH spin distribution peaks at high ($\sim 0.6-1$),
intermediate ($\sim 0.4$), and low values ($\sim 0.1$) for those
cases with $f_\tde=0.1$, $0.5$, and $0.9$, respectively, while the 
MBH spins cover a broad range from $0$ to $1$ and the distribution 
is roughly flat from $0.2$ to $0.6$ when $f_\tde$ is randomly 
distributed between $0.1$ and $0.9$. If $M_\cl=10^5 \msun$ (right 
panels), the spin increase during a single AGN episode is more 
significant than that for $M_\cl = 10^4\msun$. Therefore, the resulting 
spin distributions shift towards higher spin values compared with those 
corresponding cases of $M_\cl = 10^4\msun$, although the accretion 
of TDSs is still efficient in slowing the rotation of the central MBH. In 
the case of $M_\cl = 10^5\msun$, the spin distribution peaks at a value 
close to $1$ with a skewed wing towards $0$ when $f_\tde = 0.1$, 
$0.5$, or randomly distributed between $0.1$ and $0.9$, while the 
spins concentrate at lower values $\la 0.3$ when $f_\tde = 0.9$. The 
median spin values of those mock AGNs generated with different model 
parameters $(f_\tde,\ M_\cl)$ are listed in Table~\ref{tbl-2}, which will be 
compared with the simulation results obtained in Section~\ref{sec:implication}.

\begin{table*}
\begin{center}
  \caption{Median spin values of mock AGNs with mass $\sim 10^6M_{\odot}$ predicted from our models
   and the effective spins obtained from the stacked `NLS1' X-ray spectra assuming the model spin distributions.}
	\label{tbl-2}
	\begin{tabular}{ccccccc}
    \tableline
     \multirow{2}{*}{$f_{\rm TDE}$}    &   \multirow{2}{*}{$M_{\rm cl} (\msun)$}  & \multicolumn{2}{c}{Without SG }
     & & \multicolumn{2}{c}{With SG} \\ \cline{3-4} \cline{6-7} &     &  med-Spin   & eff-Spin   &  & med-Spin  & eff-Spin \\
    \tableline
    0.1             &       $10^4$       &      $0.83^{+0.11}_{-0.15}$   &      $0.67^{+0.27}_{-0.49}$  	&  &     $0.80^{+0.09}_{-0.13}$   &      $0.77^{+0.20}_{-0.40}$ \\
    0.5             &       $10^4$       &      $0.44^{+0.13}_{-0.09}$   &      $0.50^{+0.35}_{-0.41}$  	&  &    $0.44^{+0.06}_{-0.06}$   &      $0.31^{+0.33}_{-0.31}$  \\
    0.9             &       $10^4$       &      $0.11^{+0.05}_{-0.04}$   &      $0.27^{+0.40}_{-0.27}$  	&  &    $0.11^{+0.04}_{-0.03}$   &        $0.21^{+0.34}_{-0.21}$\\
    ran($0.1$-$0.9$)      &       $10^4$       &      $0.44^{+0.28}_{-0.26}$   &      $0.50^{+0.34}_{-0.40}$  	&  &    $0.42^{+0.23}_{-0.25}$   &      $0.37^{+0.37}_{-0.22}$  \\
    0.1             &       $10^5$       &      $0.99^{+0.01}_{-0.05}$   &      $0.71^{+0.29}_{-0.46}$  	&  &    $0.82^{+0.07}_{-0.08}$   &      $0.69^{+0.22}_{-0.34}$  \\
    0.5             &       $10^5$       &      $0.68^{+0.28}_{-0.25}$   &      $0.60^{+0.31}_{-0.49}$  	&   &   $0.46^{+0.08}_{-0.14}$   &      $0.43^{+0.21}_{-0.34}$  \\    
    0.9             &       $10^5$       &      $0.20^{+0.21}_{-0.11}$   &      $0.21^{+0.40}_{-0.21}$  	&   &   $0.11^{+0.05}_{-0.05}$   &       $0.23^{+0.30}_{-0.23}$ \\
    ran($0.1$-$0.9$)      &       $10^5$       &      $0.72^{+0.27}_{-0.40}$   &      $0.53^{+0.32}_{-0.50}$  	&   &   $0.46^{+0.24}_{-0.25}$   &      $0.50^{+0.32}_{-0.34}$  \\
    0.1             &       ran($10^4$-$10^5$)       &      $\cdots\cdots$				      &      $\cdots\cdots$					&     & $0.80^{+0.08}_{-0.11}$   &      $0.71^{+0.25}_{-0.30}$ \\
    0.5             &       ran($10^4$-$10^5$)       &     $\cdots\cdots$				      &      $\cdots\cdots$					&     & $0.45^{+0.09}_{-0.10}$   &      $0.49^{+0.22}_{-0.30}$ \\    
    0.9             &       ran($10^4$-$10^5$)       &     $\cdots\cdots$				      &      $\cdots\cdots$					&     & $0.12^{+0.05}_{-0.05}$   &       $0.21^{+0.28}_{-0.21}$\\
    ran($0.1$-$0.9$)      &       ran($10^4$-$10^5$)       &      $\cdots\cdots$			      &      $\cdots\cdots$					&    &  $0.50^{+0.22}_{-0.26}$   &      $0.40^{+0.35}_{-0.40}$ \\
    \tableline
    \end{tabular}
\end{center}
\tablecomments{The first two columns list the model parameters $f_\tde$ 
and $M_\cl$, respectively. The third column lists the median values of the 
spin distributions resulting from each model (in terms of $f_{\rm TDE}$ and 
$M_{\rm cl}$) with $M_\disk=M_\cl$, without considering of disk self-gravity 
and fragmentation. The fourth column lists the effective spin obtained from 
the simulation for the stacked X-ray spectrum of $51$ mock NLS1s similar 
to that in \citet{Liuetal15} by assuming that NLS1s have relativistically 
broadened Fe K$\alpha$ emission and the spin distribution of those NLS1s 
are the same as that from our model in the third column (see 
Section~\ref{sec:implication}). The fifth and sixth columns list the median 
and effective spin values resulting from the models with the same $f_{\rm 
TDE}$ and $M_{\rm cl}$ as those for the third and fourth columns, but 
considering the disk self-gravity and fragmentation in the model (i.e., 
$M_\disk = \min(M_\cl, M_\sg$)). The superscript and subscript numbers 
associated with each spin value represent the $16$th and $84$th percentiles. }
\end{table*}

According to the comparison between the left and right panels of 
Figure~\ref{fig-pa}, choosing an $M_{\rm cl}$ substantially larger 
than $10^5\msun$, which means only one or a few gas-cloud 
(AGN) accretion episodes, will lead to the spins more concentrated 
at higher values for $f_\tde=0.1$, $0.5$, or $f_\tde$ randomly 
distributed over the range from $0.1$ to $0.9$, and the peak of 
spin distribution resulting from the case with $f_\tde =0.9$ shifts 
towards an even higher value.

For comparison, we plot Figure~\ref{fig-pa_sg} to show the results 
for those cases with disk mass capped by the self-gravity constraint, 
considering a constant gas-cloud mass of $10^4\msun$ or $10^5\msun$ 
for all MBHs, or a gas-cloud mass randomly selected in the range 
of $10^4$-$10^5\msun$ for each AGN accretion episode. All of these 
cases show similar spin distribution as that for $M_\disk=M_\cl=10^4
\msun$ (left panels in Fig.~\ref{fig-pa}). This is because $M_\sg$ is 
comparable to $M_\cl$ when $M_\bh\sim 10^6\msun$ as mentioned 
above to explain Figure~\ref{fig-am}. If considering disk self-gravity, 
a gas-cloud with mass much larger than $10^4 \msun$ forms a disk 
of $\sim 10^4\msun$. This is the reason why the spin distribution 
indicated by the thick red dotted-line in the right panel of 
Figure~\ref{fig-pa_sg} does not peak at $\sim1$, apparently different 
from the one without considering the disk self-gravity (thin red dotted-line). 

\begin{figure*} 
\centering
\includegraphics[scale=0.85]{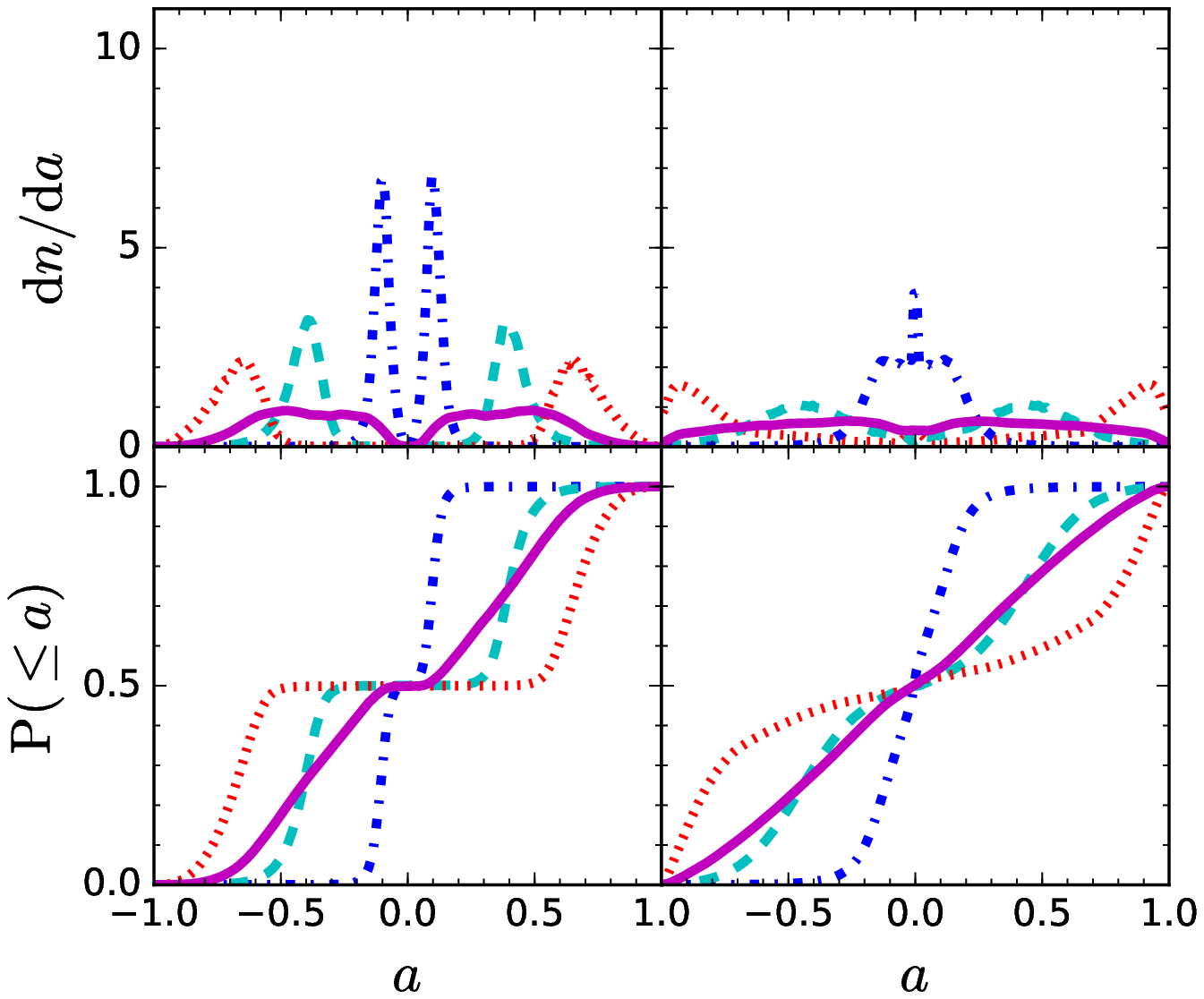}
\caption{Differential (upper panels) and cumulative (lower panels)  
spin distributions of those mock MBHs ``detected'' in the TDS accretion 
phases obtained from simulations for different models. Each of
the lines in both the left and right panels represents for a model
with $M_\cl$ and $f_\tde$ the same as that shown in Fig.~\ref{fig-pa}.
}
\label{fig-tde}
\end{figure*}

Note that in our models we do not set gas-cloud mass $\lesssim 
10^3\msun$ as observations hint larger disk mass (see discussion 
in Section~\ref{sec:subsecchao}). However, if the gas-cloud mass 
is smaller than $10^3\msun$ and the disk-to-MBH mass ratio is 
$\lesssim 10^{-3}$ for $\sim 10^6\msun$ MBHs, then MBHs cannot 
be spun up to large values (e.g., $\lesssim0.4$; see the analytic
analysis in \citealt{KPH08} and Fig.~11 in \citealt{zhanglu19}) either 
in a single gas-cloud accretion episode or after a large number of 
chaotic accretion episodes, and this will result in a spin distribution 
oscillating around a small value with a scatter determined by the 
disk-to-MBH mass ratio and $f_\tde$ \citep[e.g., see Figs. 3 and 4 
in][]{KPH08}.  Our calculations further show that significant TDS 
accretion would lead to not only smaller spins but also a narrower 
spin distribution for a given disk-to-MBH mass ratio, i.e., the larger 
the $f_\tde$, the smaller the spin values and its scatter, which is 
basically the difference that may be used to distinguish TDS accretion 
from small gas-cloud (with mass $\gtrsim 10^{3}\msun$) accretion 
(e.g., Figs.~\ref{fig-pa} and \ref{fig-tde}).

\begin{figure*} 
\centering
\includegraphics[scale=0.85]{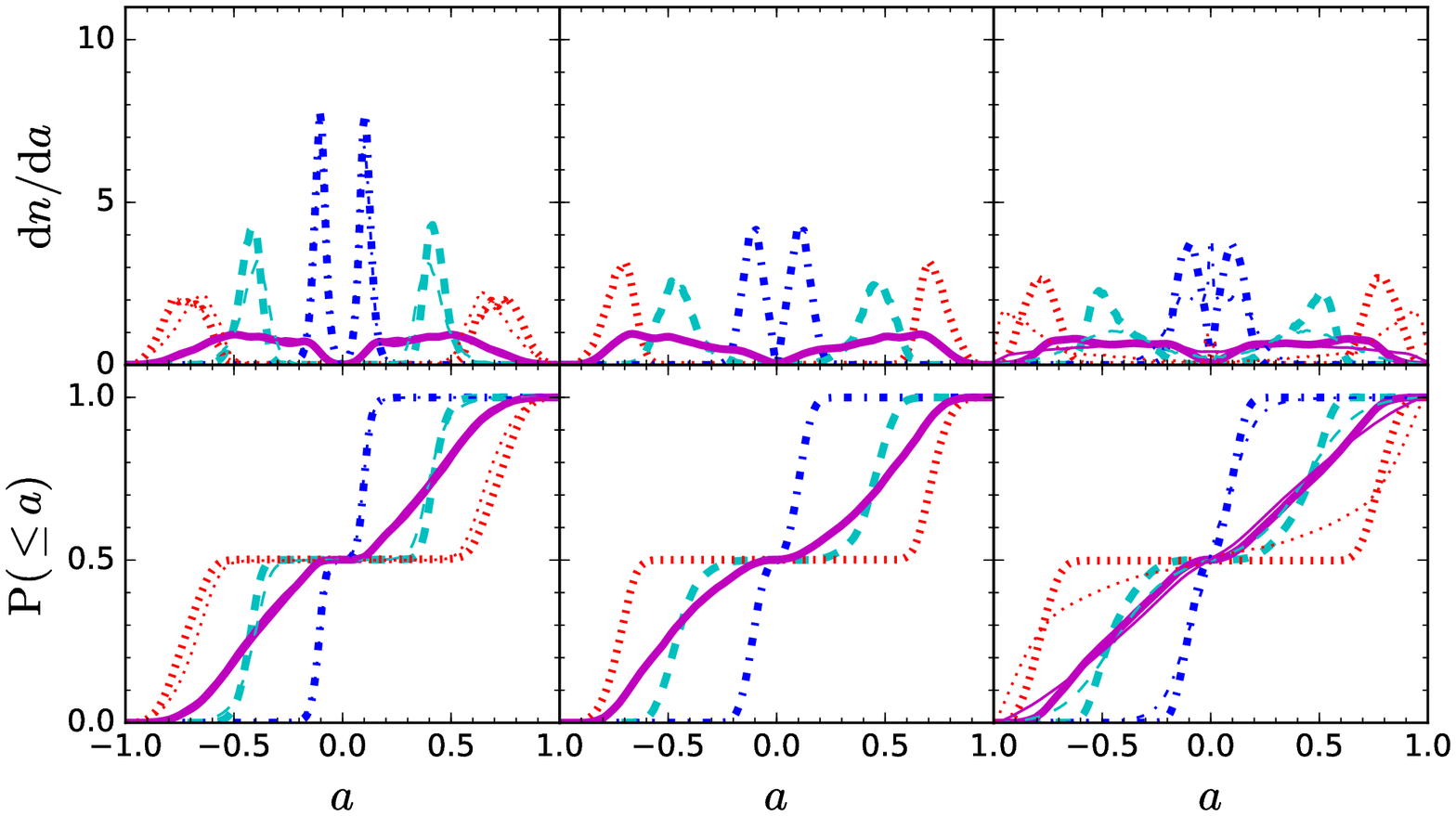}
\caption{Legend similar to Figure~\ref{fig-tde}. 
Thick curves in each
panel represent the results obtained from those models by considering
the self-gravity regulation of the disks in the gas-cloud accretion
episodes, i.e., $M_\disk=\min(M_{\rm sg},\ M_\cl)$ (see
Table~\ref{tbl-2}). Left and right panels show the cases with
assumption of
$M_\cl=10^4 \msun$ and $10^5 \msun$, respectively; middle panels show
the cases with assumption of $M_\cl$ randomly distributed over $10^4 \msun$ and
$10^5 \msun$. For comparison, the results shown in Fig.~\ref{fig-tde}
for models without considering the self-gravity regulation of the
disks are also correspondingly shown here by thin curves in the left
and right panels.
}
\label{fig-tde_sg}
\end{figure*}

Figure~\ref{fig-tde} shows differential and cumulative distributions 
of spins for those mock samples of MBHs selected in the TDS accretion 
phases, also according to the mass and spin evolutionary curves obtained 
from our simulations by setting different $f_\tde$ and $M_\cl$ without 
consideration of disk self-gravity. 
Now the prograde and retrograde spins are almost symmetrically distributed. 
The reason is that the alignment timescale is short and the probability for 
anti-alignment is almost the same as that for alignment because $\left| {\bf 
J}_{\rm disk} \right|$ is much smaller than $\left| {\bf J}_{\bullet} \right|$.
For those cases with $M_\cl = 10^4\msun$ (left panels), the spin distribution 
peaks at $\pm 0.1$, $\pm 0.4$, and $\pm 0.7$ when $f_\tde = 0.9$, $0.5$, 
and $0.1$, respectively, while it covers a broad range from -1 to $1$ and the 
distribution is rather flat in the range of $0.2<|a|<0.6$ when $f_\tde$ is randomly 
distributed between $0.1$ and $0.9$. For those cases with $M_\cl = 10^5\msun$ 
(right panels), the spins $|a|$ are more concentrated at high ($\ga 0.8$) or low
values ($\la 0.2$) when $f_\tde = 0.1$ or $0.9$, while it is close to a flat distribution 
over the range from 0 to $1$ when $f_\tde=0.5$ or $f_\tde$ is randomly set over 
the range from $0.1$ to $0.9$ for MBHs.
By comparing the same type of lines shown in Figures~\ref{fig-pa} and 
\ref{fig-tde} and thus the same model parameters $(f_\tde, M_{\rm cl})$, 
we can see that MBHs ``detected'' at the gas-cloud (AGN) accretion 
episodes have relatively high spins compared with those MBHs detected 
in the TDS accretion phase for $f_\tde \ga 0.5$. For example, for the case 
of $M_\cl=10^4 \msun$ and $f_\tde=0.1\ (0.5)$, half of those mock MBHs 
in the gas-cloud (AGN) accretion episodes have spins $|a|>0.83\ (0.45)$, 
while only about $7\%\ (28\%)$ of mock MBHs in the TDS accretion phase
have spins $|a|>0.83\ (0.45)$ (see the red dotted and cyan dashed lines in 
the left-bottom panel of the two figures). 
The reason is that the decrease of MBH spin during the period of consecutive
TDS accretion is more rapid than the increase of MBH spin during the AGN 
period when the spin is large (see Fig.~\ref{fig-am}).

Similarly, considering the self-gravity of disks during the AGN episodes 
results in similar spin distribution (see Figure~\ref{fig-tde_sg}) for MBHs 
``detected'' in TDE phase as that for $M_\disk=M_\cl=10^4 \msun$ 
(left panels in Figure~\ref{fig-tde}. Choosing an $M_\cl$ much larger 
than $10^4\msun$ does not affect the spin distribution when considering
self-gravitating disks.  

To close this section, we note here that the MBH accretion history 
assumed in our models is described by only two parameters, which 
may be too simple to reflect the reality. However, the quantitatively 
investigation presented in the present paper on how the spin evolution 
is affected by the amount of mass contributed by TDSs may provide 
a first step for future more detailed and more realistic studies. With 
such a framework, future measurements of MBH spins may be used 
to put constraints on the significance of TDSs accretion to $\sim 10^6
\msun$ MBHs and also properties of gas-clouds in the AGN accretion 
episodes. 

\section{Implications from current spin measurements}
\label{sec:implication}

The spins of more than two dozen MBHs have been estimated through 
the profile of relativistic Fe K$\alpha$ lines detected in their X-ray
spectra. According to \citet{Brenneman13} and \citet{Reynolds14},
six MBHs with spin measurements from the Fe K$\alpha$ line have 
masses in the range of $10^6-10^{6.5}\msun$.  Most of these MBHs 
have high spins, i.e.,  $a >0.99$, $>0.98$, $>0.97$, $0.96^{+0.01}_{-0.07}$, 
and $\ge0.84$ for NGC\,4051, MCG 6-30-15, 1H 0707-495, Ark 564, 
and NGC 1365, respectively. Only the small MBH ($\sim 1.1\times 
10^6\msun$) in Mrk 359 may have a relatively small spin of 
$0.66^{+0.30}_{-0.54}$ (see Table~\ref{tbl-1}). The cumulative distribution 
of those spin measurements is also shown in the bottom panels of 
Figure~\ref{fig-pa} and \ref{fig-pa_sg} (the black curve). To obtain this cumulative 
spin distribution, we have taken account of the error budgets of each 
spin measurement listed in Table~\ref{tbl-1}. For each of the four objects
with lower limits at $90\%$ confidence level, we assume its spin has $90\%$
probability to distribute uniformly in between the lower limit and the maximum 
spin value $0.998$, and $10\%$ probability to distribute uniformly in between 
$-0.998$ and the lower limit. For each of the two spin measurements with 
asymmetric errors, we assume two half-Gaussians for the spin probability 
distribution, with the two error values the dispersions of the two Gaussian 
distributions, and we also adopt the cut $|a|\leq 0.998$. A consensus on the 
reliability of these spin measurements may still not be reached 
\citep[e.g.,][]{Done16}, but if these measurements are accurate as believed 
and if there is no significant selection bias, then it may suggest that the
contribution from TDS accretion to the growth of $10^6-10^{6.5}\msun$ 
MBHs is not very significant ($\la 10\%$; see the bottom panels of 
Fig.~\ref{fig-pa} and \ref{fig-pa_sg}).

However, the Fe K$\alpha$ line profile measured from the stacked 
X-ray spectrum of 51 NLS1s indicates that the effective spin value 
of the MBHs in those NLS1s must be $<0.81$ and may be even 
$< 0.66$ (at $90\%$ confidence level) as shown in \citet{Liuetal15}. 
The two open circles with left arrows marked in Figure~\ref{fig-pa} and \ref{fig-pa_sg})
show such constraints. As seen from Figure~\ref{fig-pa}, apparently,
the low effective spin may suggest that the contribution from the 
accretion of TDSs to the MBH growth must be comparable to or even 
more significant than that from the gas-cloud (AGN) accretion episodes.

To make this comparison more clear, we further simulate X-ray 
spectra of $51$ mock AGNs similar to that in \citet{Liuetal15}. 
The X-ray spectral model used in our simulations consists of 
an absorbed power-law continuum and a broad Fe K$\alpha$ line. 
The parameters of the continuum model for each mock AGN are 
taken from the best-fitting result of \citet{Liuetal15} for each of 
those NLS1s. The broad Fe K$\alpha$ line emission is generated 
by assuming a spin randomly drawn from the spin distribution 
resulting from each of the $20$ models listed in Table~\ref{tbl-2} 
(see also Fig.~\ref{fig-pa}and \ref{fig-pa_sg})). The inclination of the disk in the 
broad-line model is fixed at $30$\arcdeg. The equivalent widths 
(EWs) of the broad lines are randomly drawn from a probability 
distribution constructed from the measured EW values for the two 
dozen sources with spin measurements listed in \citet[][two sources 
with extremely strong broad Fe K$\alpha$ line, i.e., $EW > 600$\,eV, 
are excluded]{Brenneman13}. Then we obtain a stacked spectrum 
by stacking those $51$ simulated X-ray spectra for each of the $20$ 
models. The X-ray data for the majority of the mock AGNs are 
dominated by statistical uncertainties (low signal-to-noise ratio). 
In order to estimate the effective spin value as well as its uncertainty,  
we repeat the above simulation process, and generate $100$ stacked 
spectra for each spin distribution. We estimate the MBH spin values 
(the typical $1\sigma$ uncertainty is about $\sim0.3$) by fitting each 
of the $100$ stacked spectra with an absorbed power-law continuum 
plus a relativistic broad Fe K$\alpha$ model. The median of the $100$ 
measured spin values is then considered as the ``best-fit" effective spin 
value estimated from the stacked X-ray spectra for each of the $20$ spin 
distributions resulting from those models in Section~\ref{sec:results}. 
The lower and upper limits of the spin values are estimated using the 
$16$-th and $84$-th percentiles (see Table \ref{tbl-2}).

For the cases without consideration of the disk self-gravity in the chaotic 
accretion phase, our calculations suggest that it is possible to distinguish 
the model with $(f_\mathrm{TDE}, M_\mathrm{cl}) =(0.9, 10^4\msun)$  
from that with $(f_\mathrm{TDE}, M_\mathrm{cl}) =(0.1, 10^4\msun)$ 
using the X-ray spectral stacking method. The former results in a 
median spin value of \textbf{$a_{\rm med}\sim0.27$} while the latter 
results in a much larger value of \textbf{$a_{\rm med} \sim 0.67$}. 
This is also true for those models with \textbf{$M_\mathrm{cl}=10^5
\msun$}, i.e., \textbf{$a_{\rm med} \sim0.21$} for $f_\mathrm{TDE}
=0.9$ and $a_{\rm med}\sim 0.71$ for $f_\mathrm{TDE}=0.1$. 
However, we cannot distinguish the model with $f_\mathrm{TDE}=
0.5$ from that with $f_\mathrm{TDE}$ uniformly distributed over 
$0.1$ to $0.9$, i.e., the measured MBH spins are $a_{\rm med} 
\sim0.5-0.6$ for both models. 
For the cases considering self-gravitating disks in the chaotic phase and 
the three different choices of $M_\cl$ (the last column of Table~\ref{tbl-2}), 
the models with $f_{\rm TDE}=0.1$ and $0.9$ can also be distinguished, 
which result in a median spin value of $\sim 0.7$ and $\sim 0.2$ respectively.
Similarly, the models with $f_{\rm TDE}=0.5$ and those with randomly distributed 
$f_{\rm TDE}$ cannot be distinguished within $20\%$ uncertainty. Models with
different choices of $M_\cl$ cannot be distinguished since the disk self-gravity
plays a major role in determining the disk mass and thus the median spin.
Consideration of disk self-gravity in general results in a smaller median spin, 
which is consistent with the results presented in Section~\ref{sec:results}. 
However, the two cases with $(f_\mathrm{TDE}, M_\mathrm{cl}) 
=(0.1, 10^4\msun)$ and $(0.9, 10^5\msun)$  by considering the disk self-gravity result in slightly larger spins than those cases without considering disk self-gravity (but they are consistent with each other  within the uncertainties), which may be caused by the low 
signal-to-noise ratio of the simulated spectra.
According to those simulation results, 
the observational results obtained by \citet{Liuetal15} perhaps suggest 
that those models with $f_\mathrm{TDE} \ga 0.5$ or $f_\mathrm{TDE}$ 
uniformly distributed over the range from $0.1$ to $0.9$ are favored, 
i.e., the accretion of TDSs contributes significantly to the mass growth 
of $\sim10^6\,\msun$ MBHs.

Apparently the constraints on $f_\mathrm{TDE}$ obtained from the 
two different sets of observational results on the MBH spins are in 
contradiction with each other. The reason for these inconsistent 
constraints on $f_\mathrm{TDE}$ might be that (1) the measurements 
of spins for those objects listed in Table~\ref{tbl-1} may be biased 
towards high values, or (2) a significant fraction of those NLS1s do 
not have relativistic Fe K$\alpha$ emission from the inner region 
of their disks, which leads to the immerse of the red wing of Fe K$\alpha$ 
line in the stacked continuum and thus an underestimate of the effective 
spin. Future measurements of spins for a large unbiased sample of AGNs 
would be important in constraining the significance of the contribution from 
accretion of TDSs to the growth of MBHs with mass $\sim 10^6\msun$.

\section{Conclusions and Discussions}
\label{sec:conclusions}

In this paper, we have quantitatively investigated the effect of
accreting TDSs on the spin evolution of MBHs with mass $\sim
10^6\msun$, by considering accretion of both TDSs and gas 
clouds (with several or many chaotic AGN accretion episodes). 
We find that the accretion of TDSs may play an important or
even a dominant role in shaping the spin distribution of $\sim
10^6\msun$ MBHs, depending on the contribution fraction of 
the TDS accretion to the MBH growth, which is longly expected
to be significant. Assuming a reasonable range for the mass of 
gas clouds in AGN accretion episodes, we find that most $\sim 1
0^6\msun$ MBHs in both the gas-cloud (AGN) and TDS accretion 
episodes would have: (1) low spins ($|a|\la 0.3$) if the contribution 
of TDS accretion to the growth of those MBHs is $\ga 90\%$; (2) 
high spins ($|a|\ga 0.7$) if this contribution is $\la 10\%$; and (3) 
intermediate spins or $|a|$ widely distributed over the range from
$0$ to $1$ if this contribution is intermediate or randomly distributed 
from $10\%$ to $90\%$ among those MBHs. We also find that there 
are fewer high-spin MBHs in the TDS accretion episodes than those
in the gas-cloud accretion (AGN) episodes. This asymmetry simply 
results from that the decrease of MBH spins during the period of 
consecutive TDS accretion is more rapid than the increase of MBH 
spins during the AGN period when those spins are large. One should 
also note that the occurrence of retrograde accretion is almost the 
same as prograde one in the TDS accretion episodes, while the 
fraction of retrograde-spin one in the gas-cloud accretion (AGN) 
episodes is negligible.

By comparing our model results on the spin distribution with the
observational measurements of MBH spins via Fe K$\alpha$ line 
for a number of individual AGNs with mass $\sim 10^6\msun$, we 
find that the contribution to MBH growth from accreting TDSs is 
insignificant ($\la 10\%$). However, the constraint on spins 
obtained from the stacked X-ray spectra of a larger number of 
NLS1s \citep{Liuetal15} suggests the contribution of TDS accretion 
to the growth of MBHs with mass $\sim 10^6\msun$ may be 
significant ($\ga 50\%$), which is in contradiction with the constraint 
inferred from the individual spin measurements. A large unbiased 
sample of spin measurements for AGNs is required in order to put a  
strong constraint on the contribution of accretion of TDSs to
the growth of MBHs with mass $\sim 10^6\msun$.

Current observations of TDEs are still not able to put constraints on 
the spins of their central MBHs. However, it is expected that future 
observations can provide strong constraints on the MBH spins through 
spin-induced Lense-Thirring precession \citep[][]{sto12}, Quasi-Periodic 
Oscillations (QPOs) \citep{Reis12, Pasham19}, or other signatures. With 
QPO measurement in ASASSB-14li, a TDE with central MBH mass $\sim 
10^{5.8-7.1}\msun$, \citet{Pasham19} recently found that the spin of this 
MBH should be greater than $0.7$. This high spin in the TDS accretion 
state may also suggest that the contribution of TDS accretion to MBH 
growth is not significant (see Fig.~\ref{fig-am_tde}), similar to that by the 
Fe K$\alpha$ line measurements of spins for other $\sim 10^6\msun$ 
MBHs in the AGN states (see Table~\ref{tbl-1}).

We anticipate that future measurements on the spins of some $\sim
10^6\msun$ MBHs and determination of their spin distribution in both
the gas-cloud (AGN) and the TDS accretion episodes will put strong
constraints on the significance of the accretion of TDSs to the growth 
history of those MBHs. In addition, the comparison of the two spin 
distributions would put further constraints. For example, the accretion 
of TDSs to the growth of $\sim 10^6\msun$ MBHs would be important  
if the fraction of highly spinning MBHs in the TDS accretion episodes 
is found to be relatively small comparing to that in the gas-cloud (AGN) 
accretion episodes.

\acknowledgements
We thank Qingjuan Yu for helpful discussions on and contributions to 
various aspects presented in this paper.
This work is partly supported by the National Key Program for Science
and Technology Research and Development (Grant No. 2016YFA0400704), 
the National Natural Science Foundation of China (Grant No. 11873056,
11690024,  and  11390372), and the Strategic Priority Program of the 
Chinese Academy of Sciences (Grant No. XDB 23040100).



\begin{thebibliography}{}
%
\bibitem[Artymowicz et al.(1993)]{Art93} Artymowicz, P., Lin, D.~N.~C., 
\& Wampler, E.~J.\ 1993, \apj, 409, 592 
%
\bibitem[Bardeen \& Petterson(1975)]{bar75} Bardeen, J. M., \&
Petterson, J. A. 1975, \apj, 195, 65
%
\bibitem[Barausse(2012)]{Barausse12} Barausse, E.\ 2012, \mnras, 423, 2533 
%
\bibitem[Bardeen et al.(1972)]{Bardeen72} Bardeen, J.~M., Press,
W.~H., \& Teukolsky, S.~A.\ 1972, \apj, 178, 347 
%
\bibitem[Beloborodov et al.(1992)]{Beloborodov92} Beloborodov, A.~M.,
Illarionov, A.~F., Ivanov, P.~B., \& Polnarev, A.~G.\ 1992, \mnras,
259, 209 
%
\bibitem[Berti \& Volonteri(2008)]{ber08} Berti, E., Volonteri, M.
2008, \apj, 684, 822
%
\bibitem[Blanchard et al.(2017)]{Blanchard17} Blanchard, P.~K., Nichols, M., Berger, E., et al.\ 2017, \apj, 843, 106
%
\bibitem[Brenneman(2013)]{Brenneman13} Brenneman, L. 2013, AcPol, 53,
652
%
\bibitem[Brenneman \& Reynolds(2006)]{BR06} Brenneman, L.~W., \& 
Reynolds, C.~S.\ 2006, \apj, 652, 1028 
%
\bibitem[Capellupo et al.(2016)]{2016MNRAS.460..212C} Capellupo, D.~M., Netzer, H., Lira, P., Trakhtenbrot, B., \& Mej{\'{\i}}a-Restrepo, J.\ 2016, \mnras, 460, 212 
%
\bibitem[Cui \& Yu(2014)]{CY14} Cui, X., \& Yu, Q.\ 2014, \mnras, 437, 777 
%
\bibitem[de Felice(1980)]{defelice80} de Felice, F.\ 1980, Journal of
Physics A Mathematical General, 13, 1701 
%
\bibitem[Done \& Jin(2016)]{Done16} Done, C., \& Jin, C.\ 2016, \mnras, 460, 1716 
%
\bibitem[Donley et al.(2002)]{donley02} Donley, J. L., Brandt, W. N., Eracleous, M., 
Boller, Th. 2002, \aj, 124, 1308
%
\bibitem[Dotti et al.(2013)]{Dotti13} Dotti, M., Colpi, M., Pallini,
S., et al. 2013, \apj, 762, 68
%
\bibitem[Esquej et al.(2008)]{esquej08} Esquej, P., Saxton, R. D., Komossa, S., et al.
2008, \aa, 489, 543
%
\bibitem[Figer et al.(1999)]{figer99} Figer, D. F., Kim, S. S. Morris, M., et al. 1999, 
\apj, 525, 750
%
\bibitem[Foucart \& Lai(2014)]{fou14} Foucart, F., \& Lai, D. 2014,
\mnras, 445, 1731
%
\bibitem[Franchini et al.(2016)]{Franchini16} Franchini, A., Lodato,
G., \& Facchini, S.\ 2016, \mnras, 455, 1946 
%
\bibitem[Frolov \& Novikov(1998)]{Frolov98} Frolov, V.~P., \& Novikov,
I.~D.\ 1998, Black hole physics : basic concepts and new developments
/by Valeri P.~Frolov and Igor D.~Novikov.~Dordrecht : Kluwer Academic,
c1998.~(Complete rewriting of The Physics of Black Holes, 1989,
Kluwer.)
%
\bibitem[Gezari et al.(2008)]{gazari08} Gezari, S., Basa, S., Martin, D. C.
2008, \apj, 676, 944
%
\bibitem[Goodman(2003)]{Goodman03} Goodman, J.\ 2003, \mnras, 339, 937 
%
\bibitem[Habibi et al.(2013)]{habibi13} Habibi, M., Stolte, A., Brandner, W., 
et al. \aap, 2013, 556, 26
%
\bibitem[Harfst et al.(2010)]{harfst10} Harfst, S., Portegies, Zwart, S., \&
Stolte, A. 2010, \mnras, 409, 628
%
\bibitem[Herrnstein et al.(1998)]{her98} Herrnstein, J. R., Moran, J. M., 
et al. 1998, \apj, 497, 69
%
\bibitem[Herrnstein et al.(2005)]{her05} Herrnstein, J. R., Moran, J. M., 
Greenhill, L. J. \& Trotter, A. S. 2005, \apj, 629, 719
%
\bibitem[Heyer \& Dame(2015)]{heyer15} Heyer, M., \& Dame, T. M. 2015, \araa, 53, 583 
%
\bibitem[Hills(1975)]{Hills75} Hills, J.~G.\ 1975, \nat, 254, 295 
%
\bibitem[Hughes(2000)]{Hughes00} Hughes, S.~A.\ 2000, \prd, 61, 084004 
%
\bibitem[Hughes(2001)]{Hughes01} Hughes, S.~A.\ 2001, \prd, 64, 064004 
%
\bibitem[Ishihara \& Nakai(2002)]{ishihara02} Ishhara, Y., \& Nakai, N. 2002, 
IAUS, 206, 400
%
\bibitem[Kerr(1963)]{Kerr} Kerr, R.~P.\ 1963, Physical Review Letters,
11, 237 
%
\bibitem[Kesden(2012)]{Kesden12} Kesden, M.\ 2012, \prd, 85, 024037 
%
\bibitem[King et al.(2005)]{kin05} King, A. R., Lubow, S. H., Ogilvie,
G. I., \& Pringle, J. E. 2005, \mnras, 363, 49
%
\bibitem[King \& Pringle(2006)]{KP06} King, A.~R., \& Pringle, J.~E.\
2006, \mnras, 373, L90 
%
\bibitem[King \& Pringle(2007)]{king07} King, A. R., \& Pringle, J. E. 2007, 
\mnras, 377, L25
%
\bibitem[King et al.(2008)]{KPH08} King, A.~R., Pringle, J.~E., \&
Hofmann, J.~A.\ 2008, \mnras, 385, 1621 
%
\bibitem[Kolykhalov \& Sunyaev(1980)]{kol80} Kolykhalov, P. I., \&
Sunyaev, R. A. 1980, SvAL, 6, 357
%
\bibitem[Komossa(2015)]{Komossa15} Komossa, S.\ 2015, Journal of High
Energy Astrophysics, 7, 148 
%
\bibitem[Kormendy \& Ho(2013)]{KH13} Kormendy, J., \& Ho, L.~C.\ 2013,
\araa, 51, 511 
%
\bibitem[Milosavljevi{\'c} et al.(2006)]{ML06} Milosavljevi{\'c}, M., Merritt, D., 
\& Ho, L.~C.\ 2006, \apj, 652, 120 
%
\bibitem[Lense \& Thirring(1918)]{1918PhyZ...19..156L} Lense, J., \& Thirring, H.\ 1918, Physikalische Zeitschrift, 19,  
%
\bibitem[Levin \& Beloborodov(2003)]{Levin03} Levin, Y., \&
Beloborodov, A.~M.\ 2003, \apjl, 590, L33 
%
\bibitem[Liu et al.(2015)]{Liuetal15} Liu, Z., Yuan, W., Lu, Y., \&
Zhou, X.\ 2015, \mnras, 447, 517 
%
\bibitem[Lodato \& Rossi(2011)]{Lodato11} Lodato, G., \& Rossi, E.~M.\
2011, \mnras, 410, 359 
%
\bibitem[Lu et al.(2009)]{LuJ09} Lu, J.~R., Ghez, A.~M., Hornstein,
S.~D., et al.\ 2009, \apj, 690, 1463 
%
\bibitem[Magorrian \& Tremaine(1999)]{Magorrian99} Magorrian, J., \&
Tremaine, S.\ 1999, \mnras, 309, 447 
%
\bibitem[Marconi et al.(2004)]{Marconi04} Marconi, A., Risaliti, G.,
Gilli, R., et al.\ 2004, \mnras, 351, 169 
%
\bibitem[Martine et al.(2007)]{mar07} Martin, R. G., Pringle, J. E.,
\& Tout, C. A. 2007, \mnras, 381, 1617
%
\bibitem[McHardy et al.(2005)] {mchardy05} McHardy, I. M., Gunn, K. F., Uttley, P., 
\& Goad, M. R. 2005, \mnras, 359, 1469
%
\bibitem[Menezes et al.(2018)]{menezes18} Menezes, R. B., da Silva, P., 
\& Steiner, J. E. 2018, \mnras, 473, 2198
%
\bibitem[Miralda-Escud{\'e} \& Kollmeier(2005)]{MK05} Miralda-Escud{\'e}, J., 
\& Kollmeier, J.~A.\ 2005, \apj, 619, 30 
%
\bibitem[Nayakshin \& Cuadra(2005)]{Nayakshin05} Nayakshin, S., \&
Cuadra, J.\ 2005, \aap, 437, 437 
%
\bibitem[Nelson \& Papaloizou(1999)]{nelson99} Nelson, R. P., \& Papaloizou, J. C. B. 
1999, \mnras, 309, 929
%
\bibitem[Nixon \& King(2016)] {Nixon16} Nixon, C., \& King, A. 2016, LNP, 905, 45
%
\bibitem[Novikov \& Thorne(1973)]{NT73} Novikov, I.~D., \& Thorne, K.~S.\ 
1973, Black Holes (Les Astres Occlus), 343 
%
\bibitem[Pasham et al.(2019)]{Pasham19}Pasham, D.~R., Remillard, R.~A., Fragile, P.~C., et al.\ 2019, Science, 363, 531 
%
\bibitem[Patrick et al.(2012)]{patrick12} Patrick, A. R., Reeves, J. N., Porquet, D., 
et al. 2012, \mnras, 426, 2522
%
\bibitem[Paumard et al.(2006)]{Paumard06} Paumard, T., Genzel, R.,
Martins, F., et al.\ 2006, \apj, 643, 1011 
%
\bibitem[Perego et al.(2009)]{per09} Perego, A., Dotti, M., Colpi, M.,
\& Volonteri, M. 2009, \mnras, 399, 2249
%
\bibitem[Peterson et al.(2004)] {peterson04} Peterson, B. M., Ferrarese,
L., Gilbert, K. M., et al. 2004, \apj, 613, 682 
%
\bibitem[Rees(1988)]{Rees88} Rees, M.~J.\ 1988, \nat, 333, 523 
%
\bibitem[Reis et al.(2012)]{Reis12} Reis, R.~C., Miller, J.~M.,
Reynolds, M.~T., et al.\ 2012, Science, 337, 949 
%
\bibitem[Reynolds(2014)]{Reynolds14} Reynolds, C. S. 2014, SSRv, 183,
277
%
\bibitem[Risaliti et al.(2013)] {risaliti13} Risaliti, G., Harrison, F. A., Madsen, K. K., et al.
2013, \nat, 494, 449
%
\bibitem[Risaliti et al.(2009)]{risaliti09} Risaliti, G., Miniutti, G., Elvis, M., et al. 
2009, \apj, 696, 160
%
\bibitem[S{\k a}dowski et al.(2011)]{Sadowski11} S{\k a}dowski, A.,
Bursa, M., Abramowicz, M., et al.\ 2011, \aap, 532, A41 
%
\bibitem[Shakura \& Sunyaev(1973)]{sha73} Shakura, N. I., \& Sunyaev,
R. A. 1973, \aap, 24, 337
%
\bibitem[Shankar et al.(2009)]{Shankar09} Shankar, F., Weinberg,
D.~H., \& Miralda-Escud{\'e}, J.\ 2009, \apj, 690, 20 
%
\bibitem[Sesana et al.(2014)] {ses14} Sesana, A., Barausse, E., Dotti,
M., Rossi, E. M. 2014, \apj, 794, 104
%
%
\bibitem[So\l{}tan(1982)]{Soltan82} So\l{}tan, A.\ 1982, \mnras, 200, 115 
%
\bibitem[Stone \& Loeb(2012)]{sto12} Stone, N., \& Loeb, A. 2012,
Phys. Rev. Lett., 108, 061302
%
\bibitem[Stone et al.(2013)]{Stone13} Stone, N., Loeb, A., \& Berger, E.\ 
2013, \prd, 87, 084053 
%
\bibitem[Stone \& Metzger(2016)]{stone16} Stone, N. C., \& Metzger, B. D., 
2016, \mnras, 455, 859
%
\bibitem[Strubbe \& Quataert(2009)] {strubbe09} Strubbe, L. E., \& 
Quataert, E. 2009, \mnras, 400, 2070
%
\bibitem[Thorne(1974)]{Thorne74} Thorne, K.~S.\ 1974, \apj, 191, 507 
%
\bibitem[Trakhtenbrot(2014)]{2014ApJ...789L...9T} Trakhtenbrot, B.\ 2014, \apjl, 789, L9 
%
\bibitem[van Velzen \& Farrar(2014)]{vanVelzen14} van Valzen, S., \&
Farrar, G. R. 2014, \apj, 792, 53
%
\bibitem[Vasiliev et al.(2015)]{Vasiliev15} Vasiliev, E., Antonini, F., \& Merritt, 
D.\ 2015, \apj, 810, 49 
%
\bibitem[Vasudevan et al.(2016)] {vas16} Vasudevan, R. V., Fabian, A.
C., Reynolds, C. S., et al. 2016, \mnras, 458, 2012
%
\bibitem[Volonteri et al.(2005)]{vol05} Volonteri, M., Madau, P.,
Quataert, E., \& Rees, M. J. 2005, \apj, 620, 69
%
\bibitem[Volonteri et al.(2007)]{Volonteri07} Volonteri, M., Sikora,
M., \& Lasota, J.-P.\ 2007, \apj, 667, 704 
%
\bibitem[Volonteri et al.(2013)]{Volonteri13} Volonteri, M., Sikora,
M., Lasota, J.-P., \& Merloni, A.\ 2013, \apj, 775, 94 
%
\bibitem[Walton et al.(2013)]{walton13} Walton, D. J., Nardini, E., Fabian, A. C., 
Gallo, L. C., \& Reis, R. C. 2013, \mnras, 428, 2901
%
\bibitem[Wang \& Merritt(2004)]{WangMerritt04} Wang, J., \& Merritt,
D.\ 2004, \apj, 600, 149 
%
\bibitem[Yamauchi et al.(2012)]{yamauchi12} Yamauchi, A., Nakai, N., 
Ishihara, Y.,  Diamond, P. \& Sato, N. 2012, \pasj, 64, 103
%
\bibitem[Yu \& Tremaine(2002)]{YT02} Yu, Q., \& Tremaine, S.\ 2002,
\mnras, 335, 965 
%
\bibitem[Yu \& Lu(2004)]{YL04} Yu, Q., \& Lu, Y.\ 2004, \apj, 602, 603 
%
\bibitem[Yu \& Lu(2008)]{YL08} Yu, Q., \& Lu, Y.\ 2008, \apj, 689, 732
%
\bibitem[Zhang \& Lu(2019)]{zhanglu19} Zhang, X., \& Lu, Y. 2019, \apj, 873, 101
%
\bibitem[Zhou \& Wang(2005)]{ZW05} Zhou, X.-L., \& Wang, J.-M. 2005, \apj, 618, L83
%
\bibitem[Zoghbi et al.(2010)] {zoghbi10} Zoghbi, A., Fabian, A. C., Uttley, P., et al. 
2010, \mnras, 401, 2419
%
\end{thebibliography}
\end{document}